\documentstyle[emulateapj,epsf]{article}

\def\kms{km\,s$^{-1}$\,}
\def\vs{$V_{S}$\,\,}
\def\etal{ et~al.\rm\,}

\def\g292{G292.0+1.8}
\def\chan{{\it Chandra\,}}
\def\eins{{\it Einstein\,}}

\received{2004 June 23}
\begin{document}

\title{Exploring the Kinematics of the Oxygen-Rich Supernova Remnant \g292:
Ejecta Shells, Fast-Moving Knots and Shocked Circumstellar Material}
\submitted{Revised Version, August 13, 2005}

\author{Parviz Ghavamian\altaffilmark{1}, John P. Hughes\altaffilmark{2,3}
and T. B. Williams\altaffilmark{2,3} }

\altaffiltext{1}{Department of Physics and Astronomy, Johns Hopkins University, 3400 North
Charles Street, Baltimore, MD, 21218-2686; parviz@pha.jhu.edu}
\altaffiltext{2}{Department of Physics and Astronomy, Rutgers University, Piscataway, NJ 08854-8019;
jph@physics.rutgers.edu,williams@physics.rutgers.edu}
\altaffiltext{3}{Visiting Astronomer, Cerro Tololo Inter-American Observatory, National Optical Astronomy
Observatories.  CTIO is operated by AURA, Inc.\ under contract to the National Science
Foundation.}

\begin{abstract}

We present results of an in-depth optical study of the core collapse
supernova remnant \g292\ using the Rutgers Fabry-Perot (RFP) imaging
spectrometer.  Our observations provide a detailed picture of the
supernova remnant in the emission lines of [O~III] $\lambda$5007,
H$\alpha$ and [N~II] $\lambda$6548.  The [O~III] Fabry-Perot scans
reveal a bright crescent-shaped spur of previously known high-velocity
(V$_{radial}$\,$\sim$\,1500 \kms) O-rich ejecta located on the
eastern side of the remnant.  The spur consists of a semi-coherent structure of
mostly redshifted material, along with several clumps that have
apparently broken out of the more orderly shell-like expansion.  The
high velocity ($\gtrsim\,$600 \kms) component of the spur also displays
a scalloped morphology characteristic of Rayleigh-Taylor instabilities.
We also find a large number of fast-moving knots
(FMKs) of O-rich ejecta undetected in prior photographic plate images
and similar to features seen in Cas A.  The FMKs are distributed sparsely in
the interior of \g292\ and are seen mostly in blueshifted emission out
to $V_{radial}$\,$\approx$\,$-$1700 \kms.  The position-velocity distribution
of the FMKs can be kinematically described as a shell 3\farcm4 in radius
expanding at a velocity of 1700 \kms.  Another feature apparent in the [O~III]
scans is an equatorial belt consisting of both a bar-like structure at zero radial
velocity and a clumpy, high velocity ejecta component seen in projection along the
line of sight.  Portions of the zero-velocity bar
are spatially well correlated with a similar structure
seen in the \chan\ X-ray image of \g292.  The bar is also detected in
our H$\alpha$ RFP images at zero radial velocity, providing further
evidence that this structure is of circumstellar origin.  We find that
the optical and X-ray properties of the bar are consistent with
incomplete (partially radiative) shocks in material of moderate
densities.  There are also a number of faint,
elongated structures seen in H$\alpha$ at zero radial velocity across
the interior of \g292\ that lack [O~III] and X-ray counterparts.
These filaments may be low-density H~I clouds photoionized by hard
radiation from the interior of the remnant.   Overall these results
suggest that \g292\, is currently interacting with a low density
environment.    We find no evidence
for high velocity H$\alpha$ or [N~II] emission over the dynamical range
sampled by the RFP.  Assuming a distance of 6 kpc for \g292, we estimate a kinematic
age of (3000--3400)$\,d_{\rm 6}$ years for this remnant.

\end{abstract}

\section{INTRODUCTION}

Among the known supernova remnants (SNRs), a handful contain
fast-moving debris with optical spectra dominated by lines of oxygen
and its associated burning products.  In these metal-rich remnants,
the emission is produced by excitation of dense material from deep
within the massive progenitor star ($\gtrsim\,$10 M$_{\odot}$), laid
bare by the supernova explosion.  Dense clumps of ejecta coast along
mostly intact and undecelerated until they penetrate the hot,
overpressured gas of the reverse shock.  As shocks are driven into the
fragments they are torn apart into clumps of varying size and density.
The densest material ($n\,\gtrsim\,$10 cm$^{-3}$,
$V_{shock}\,\sim\,$100 \kms) produces optical/UV spectra dominated by
very strong oxygen line emission, with little or no emission from
lines of H, He and N commonly seen in shocked interstellar material.
The spectroscopic study of ejecta material in oxygen-rich supernova
remnants (OSNRs) provides us with a rare opportunity to study both the
kinematics of the supernova ejecta and the products of both
hydrostatic and explosive nucleosynthesis in massive stars.

To date there have been six OSNRs identified: Cassiopeia A (Minkowski
1957, Chevalier \& Kirshner 1978, 1979, Reed \etal\, 1995, Fesen
\etal\,2001), Puppis A (Winkler \& Kirshner 1985) and \g292 (Goss
\etal\, 1979, Murdin \& Clark 1979, van den Bergh 1979, Braun \etal\, 1983,
Dopita \& Sutherland 1995) in the Galaxy, 0540$-$69.3 (Kirshner
\etal\, 1989) and N132D (Lasker 1978, 1980; Morse, Winkler \& Kirshner
1995; Morse \etal\, 1996; Blair \etal\, 2000) in the LMC, E0102$-$72.2
in the SMC (Dopita, Tuohy \& Mathewson 1981, Tuohy \& Dopita 1983,
Blair \etal\, 2000) and a spatially unresolved remnant in NGC 4449
(Kirshner \& Blair 1980, Blair Kirshner \& Winkler 1983).  Although
the core-collapse supernovae which produced these OSNRs should have
all left behind rotating neutron stars (pulsars), \g292\, is the only
Galactic remnant from this class found to harbor both an active pulsar
(Camilo \etal\, 2002, Hughes \etal\, 2003) and associated pulsar wind
nebula (PWN) (Hughes \etal\, 2001).  Earlier \eins\, HRI
observations indicated that a central bar of enhanced X-ray emission
was present in \g292, superimposed on an ellipsoidal disk of emission
(Tuohy, Clark \& Burton 1982).  Recent \chan\, imagery of \g292\,
(Park \etal\, 2002, 2004; Gonzalez \& Safi-Harb 2003) have revealed that the X-ray bar is of
normal composition, suggesting that \g292\, is interacting with
circumstellar wind material.  On the basis of the rich variety of
physical processes present in \g292\, (O-rich optical and X-ray
emission, active pulsar/PWN and apparent circumstellar interaction)
one may argue that \g292\, is the best SNR for a simultaneous study of
all aspects of core collapse supernova explosions.

As with many supernova remnants known today, \g292\, was first
discovered in a radio survey (Mills, Slee \& Hill 1961).  The first
followup radio observations (Lockhart \etal\, 1977) showed a
center-filled (plerionic) morphology, leading some to conclude that
the remnant was ``Crab-like'', i.e., that synchrotron radiation from a
PWN dominated the radio appearance of the SNR.  However, further radio
observations by Braun \etal\, (1986) demonstrated that there are two
components to the radio morphology of \g292: a bright circular region
near the center approximately 4\arcmin\, across embedded in a
surrounding fainter plateau of emission approximately 8\arcmin\,
across.  In a more recent \chan\, observation of \g292, Hughes \etal\,
(2001) discovered a point source of hard X-ray emission near the
geometric center surrounded by a nonthermal halo
1\arcmin\,$-$2\arcmin\, across.  They argued that the source was a
pulsar and PWN, a conclusion which was confirmed by discovery of 135
ms radio pulsations from the point source, now designated PSR J1124$-$5916
(Camilo \etal\, 2002; Hughes \etal\, 2003).  Subsequently Gaensler \& Wallace (2003;
hereafter GW03) showed that the spectral index of the bright radio core is consistent
with synchrotron radiation from a PWN ($\alpha$\,=\,$-$0.05), while
the index of the plateau is steep ($\alpha$\,=\,$-$0.5), consistent
with particle acceleration in shocks.

X-ray emission from the
metal-rich ejecta in \g292\, was first detected in \eins\, data (Clark, Tuohy, \& Becker 1980).  Detailed
X-ray spectroscopy was performed by Hughes \& Singh (1994), who showed that the
global X-ray spectrum of \g292\, was dominated by K-shell line emission from metal-rich ejecta.
Based on abundances derived from their global spectral models, Hughes \& Singh 
(1994) argued that the ejecta composition of \g292\, matched the yields of
20 M$_{\sun}$ nucleosynthesis models best.  However, in a recent analysis of \chan\, spectra of
\g292\ Park \etal\, (2004), noting that only the X-ray lines of O, Ne, Mg and Si were detected,
suggested that the reverse shock has not yet penetrated beyond the layers of hydrostatically burned
material.  Since knowledge of the relative abundances of both hydrostatic and explosive burning products
are required for an accurate estimate of the progenitor mass from the supernova nucleosynthesis models,
Park \etal\, (2004) concluded that the progenitor mass estimate by Hughes \& Singh (1994) was no longer
reliable.  Thus the mass of the progenitor star of \g292\, remains uncertain.

The first optical imagery of \g292\, was performed by Goss \etal\, (1979), who discovered a bright, crescent-shaped
spur of nebulosity coincident with the extended radio source position.  Their spectra showed emission lines of
oxygen and neon, but little or no line emission from hydrogen, helium or nitrogen. Followup spectra of the spur acquired by Murdin
\& Clark (1979), van den Bergh (1979), Braun \etal\, (1983) and Sutherland \& Dopita (1995) showed
broad [O~III] lines extending over a velocity range $\sim\,$2000 \kms.  Narrowband optical CCD images
of \g292\, have recently been acquired by Winkler \& Long (2005; hereafter WL05).  According to all these studies,
the optical appearance of \g292\, is dominated by radiative shocks in dense supernova ejecta which have undergone negligible
mixing with the ISM.

In this paper, we report results of Rutgers Fabry-Perot (RFP) imaging
spectroscopy of \g292\, in emission lines of [O~III], H$\alpha$, and
[N~II].  Both the narrowband [O III] imagery of WL05 and our RFP scans
have revealed extensive oxygen-rich material undetected in the earlier
photographs, including previously undetected knots of O-rich ejecta
similar to those seen in Cas A and an equatorial belt of shocked
material.  As we will show, the compact knots of O-rich ejecta exhibit
large radial velocities ($\sim$1500 \kms), showing that these features
are fast-moving knots (FMKs).  In addition, we will show that the zero
velocity [O III] and H$\alpha$ emission from the equatorial belt trace
a bar-like structure matching a similar feature in the X-ray band.  This
confirms the assertions of Park \etal\, (2002, 2004) and
Gonzalez \& Safi-Harb (2003) that this feature is comprised of
circumstellar material. The velocity range of both the FMKs and the
bright O-rich spur extends over a 50\% broader range in velocities
than previously determined for \g292\, (Murdin \& Clark 1979, van den
Bergh 1979), extending to $\sim$3000 \kms.  Throughout this paper
we assume a distance of 6 kpc to \g292\ (GW03) and
quote distance-dependent quantities in units of 6 kpc.

\section{OBSERVATIONS AND DATA REDUCTION}

Our observations of \g292\, were performed with the RFP spectrometer at the 1.5 m telescope of Cerro Tololo
Inter-American Observatory.  The H$\alpha$ data were acquired on the night of 1998 February 17 (UT), while
the [O~III] data were acquired on the nights of 1998 February 20 and 21 (UT).  The Tek1024 CCD and 200 mm camera
lens were used with the f/7.5 secondary focus, giving an image scale of 0\farcs65
pixel$^{-1}$ for all the observations. The field of view was a circle 7\farcm8\, in diameter, centered on 
coordinates 11$^{\rm h}$24$^{\rm m}$30\fs5, $-$59$^{\circ}$16\arcmin01\farcs3 (2000) for the H$\alpha$
scans and 11$^{\rm h}$24$^{\rm m}$24\fs2, $-$59$^{\circ}$16\arcmin08\farcs8 (2000) for the [O~III]
scans.  In Figure \ref{g292_oiii+allaps} the RFP field of view is marked on the continuum-subtracted 
narrowband [O III] image of WL05.
The broad etalon was used during all RFP observations.  The H$\alpha$ scans sampled the
emission at 19 velocities from $-$800 \kms\, to +570 \kms, at intervals of 70 \kms\, per sample.  The 
[O~III] scans were centered on 5007 \AA\, and ranged from $-$1435 \kms\, to +1675 \kms, sampling the emission from \g292\,
in 29 samples over 100 \kms\, intervals.  During our observations we isolated emission from a single spectral
order using CTIO [O~III] filter
4993/44 ($\Delta\,\lambda_{c}\,$=\,44 \AA) for scans centered in the range 4983 \AA\,$\leq\,\lambda\,\leq$5001 \AA,
filter 5007/44 for scans in the range 5003 \AA\,$\leq\,\lambda\,\leq$5019 \AA\, and 5037/44 for scans
in the range 5021 \AA\,$\leq\,\lambda\,\leq$5035 \AA.  On the first night of the [O~III] observations the
median seeing was 1\farcs9, during which we acquired the scans between $-$250 \kms\, and +710 \kms.  The
remaining FP scans were obtained during the second night of [O~III] observations, when the seeing had worsened 
to 2\farcs6.  Two 500 s frames were acquired at each RFP etalon setting in both [O~III] and H$\alpha$/[N~II].

We reduced the Fabry-Perot images using IRAF\footnote{IRAF is distributed by the National Optical Astronomy 
Observatories, which is operated by the AURA, Inc.\ under cooperative agreement with the National Science Foundation} and our
own custom software.  Each pair of FP scans was combined in IRAF using the {\sf crreject} option to
remove cosmic rays and create images with an effective exposure time of 1000 s.  We applied overscan and bias 
subtraction to all images in the standard way.  Separate
flat field images of the telescope dome white spot were obtained at each observed wavelength, and were
used to correct the detector pixel-to-pixel response variations and to remove the filter transmission function.
The wavelength scale was calibrated using a series of comparison lamp exposures of nearby H$\alpha$ and neon
emission lines.  The zero-point drift of the wavelength solution was determined from H$\alpha$ and Neon
calibration exposures interspersed with the object frames during the observations.  We estimate that
the resulting wavelength calibration is accurate to better than 0.1 \AA.
The RFP instrumental spectral profile measured from the calibration exposures is well fit by
a Voigt function with a Gaussian width of 44 \kms and Lorentzian width of 52 \kms.  The resulting instrumental
FWHM is 117 \kms.   We scaled the transmission of the RFP scans to a common airmass using the CTIO mean extinction tables.
The variations in transparency between frames was tracked via photometry of unsaturated stars in the field.  
From the photometry we calculated the multiplicative constants required to bring the fluxes from all frames to 
a common level.
Next, we applied a Gaussian smoothing kernel to each combined scan to produce a common effective seeing of 
3\farcs0\, FWHM.  Finally we computed a coordinate solution (accurate to within 0\farcs2) for each image using seven
USNO$-$A2.0 Catalogue stars distributed over the field of view.

Several factors complicate the identification of [O~III] emission features in our RFP scans.
\g292\, lies in a rich stellar field and the sizes of many emission knots are comparable to the seeing
FWHM, so there is a significant danger of mistaking stars for emission knots (and vice versa).
In addition, internal reflections within the etalon cause ghosts (mirror images)
of the brightest stars to be projected across the optical axis in each frame, further complicating the
identification of [O~III] knots. To minimize these
problems, we first subtracted the stars from each [O~III] RFP scan using a background image 
created by median combining three FP scans devoid of any obvious line emission from the SNR and the
night sky.   Scaling the background image by a multiplicative 
constant and subtracting it from each of the 28 [O~III] RFP scans, we generated a set of continuum-subtracted images.
This process removes approximately 90\% of the stars from each frame, leaving only residuals from the brightest stars 
and their ghosts.  The ghosts were identified by tracing the reflection of each bright star across the FP axis, then
removed by interpolating the surrounding sky emission over the ghost feature.  We were able to check
the completeness of our ghost removal by registering and comparing our individual FP scans with 
blue DSS images.

\section{CONSTANT VELOCITY SNAPSHOTS OF \g292 }

Much of the ejecta material in \g292 exhibits a morphology that varies strongly with velocity
and position.  To map this distribution we found it useful to extract snapshots of 
the entire remnant at a selected range of velocities.  The wavelength of the FP transmission
profile is not constant over the field but shortens quadratically from the optical axis to the
edge of each frame, amounting to a 200 \kms\, blueshift in our [O~III] observations.  Thus, 
to create the single velocity images we interpolated the flux at each point in the field using RFP scans
bracketing the desired velocity.  To minimize
source confusion we used the continuum-subtracted RFP scans to generate the constant velocity [O~III] images.  
Due to variations in sky emission between adjacent scans, the interpolation introduces a slight ringing effect
into some constant velocity images, i.e., some images show annular strips where the average pixel value
differs from the value in the adjacent region.  We applied a first order correction to the background
by adding a constant pixel value to the mismatched annular regions.  This brought the background pixel values
in all the interpolated strips to within roughly 20\% of one another. Although this procedure did not
correct for interpolation errors in the object emission and left some residual structure in the
background, it was sufficient to allow easy identification of supernova remnant emission at each velocity interval.
We were able to verify that for a given constant velocity image the brightest emission knots
do correctly appear at the velocity centroids measured from line profile fits to the knots (described
in Section~5).  To remove any further stellar residuals which may be mistaken for FMKs, we interpolated 
the background over the positions of the brightest subtracted stars in the RFP field of view.  
We examined the remaining knots visible in the image sequence of Figure \ref{g292redblue}\, and
verified that the flux from each knot followed the sequential rise and fall 
expected of real emission lines in the RFP frames.

\section{GLOBAL DISTRIBUTION OF THE O-RICH EJECTA}\label{global}

\begin{figure}
\epsscale{0.8}
\plotone{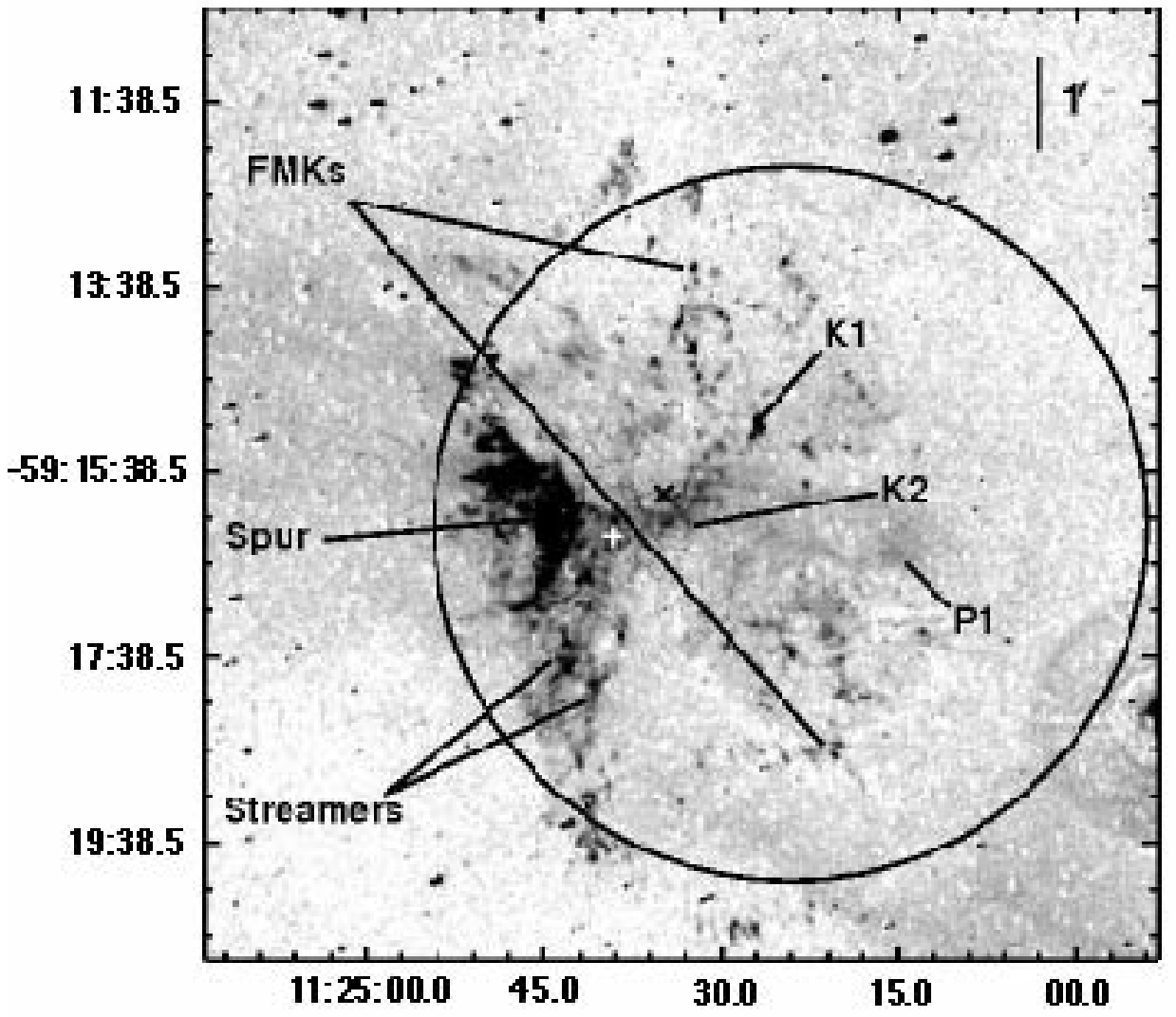}
\caption{Continuum-subtracted narrow band [O~III] image of \g292\,
(WL05), showing the optical emission sampled by the RFP observations.
The field of view of the RFP scans is marked by the solid black
circle. The radio-derived geometric center for \g292 (GW03)
is marked by the `X' and the location of the X-ray
pulsar (Hughes \etal~2003) is marked by the crosspoint.  Features
discussed in the text are marked.  N is at the top of each frame and E lies 
to the left.  A full resolution
figure can be found at http://fuse.pha.jhu.edu/\~{}parviz/papers/g292 .
}
\label{g292_oiii+allaps}
\end{figure}

\begin{figure}
\plotone{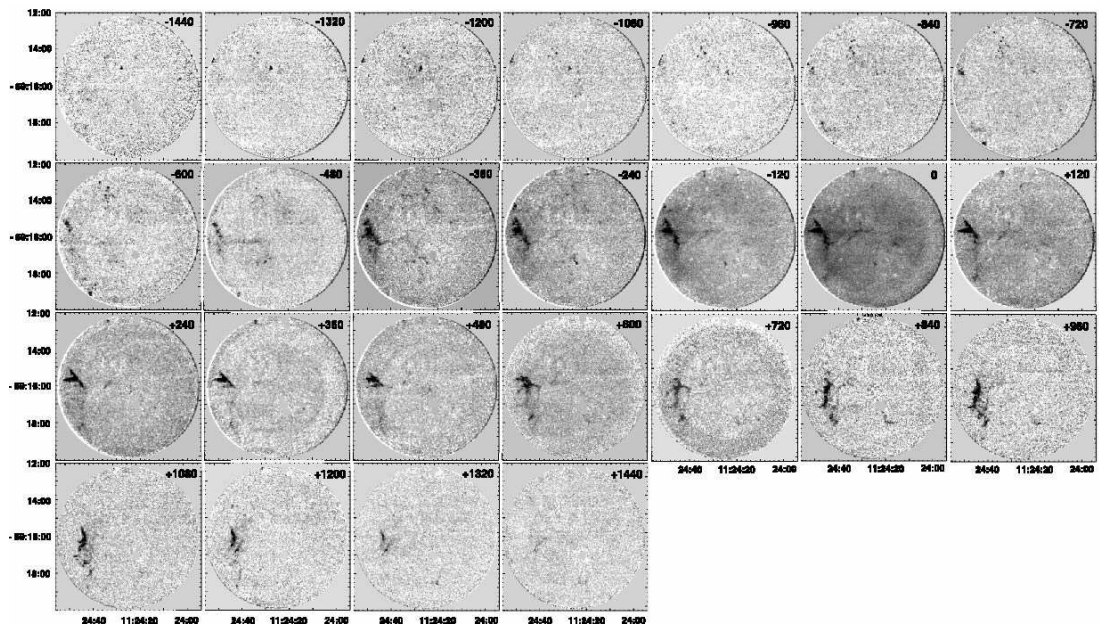}
\caption{Continuum-subtracted [O~III] scans of \g292 flattened to show
the entire SNR at a sequence of constant velocity slices.  The radial
velocity from rest is recorded in the upper right corner of each
image.  Residuals from stellar subtraction have been masked by
interpolating over the surrounding background.  The equatorial bar
and a significant portion of the eastern O-rich spur are clearly seen
in the zero velocity image, along with widespread diffuse [O~III].
Fast-moving knots (FMKs) are seen mostly in the north above the
bar.  N is at the top of each frame and E lies to the left.  
The field of view of each image is 7\farcm8.   A full resolution
figure can be found at http://fuse.pha.jhu.edu/\~{}parviz/papers/g292 .
}
\label{g292redblue}
\end{figure}

To illustrate the dependence of the global ejecta distribution on radial velocity, we present a
sequence of constant velocity [O~III] images in Figure \ref{g292redblue}.
The sequence is arranged in order of heliocentric radial velocity, from blue to red, in increments of
120 \kms\, (roughly the instrumental profile FWHM).  The sequence extends to $\pm$1440 \kms, although
there is a highly blueshifted knot near the center of \g292\, still visible blueward of $-$1440
\kms\, and a similar, but much fainter knot, detected at +1700 \kms\, (marked K1 and K2, respectively, in Figure
\ref{g292_oiii+allaps}). The faint, diffuse [O~III] emission first detected by Murdin \& Clark (1979) is
clearly visible in the image at zero radial velocity (corresponding to zero heliocentric velocity, or
V$_{LSR}\,$=\,4.5 \kms).  The diffuse
[O~III] does not appear at other velocities and extends beyond the edge of the SNR shell defined by the X-ray 
and radio observations.  This suggests that the diffuse emission is produced by photoionized interstellar gas,
in agreement with the conclusion of Murdin \& Clark (1979).

Based on the single velocity images in Figure~\ref{g292redblue}, 
the shocked [O~III] emission in \g292\, can be broadly placed into three categories:

(1) A bright, curved structure located on the eastern edge of \g292 and 
exhibiting a broad velocity dispersion ($\gtrsim$\, 2000 \kms; Murdin \& Clark 1979).  Spectroscopic observations
of this spur showed only emission lines of O and Ne (Murdin \& Clark 1979, van den Bergh 1979, Braun \etal\, 1983, Sutherland
\& Dopita 1995), indicating that the material was stellar ejecta uncontaminated by interstellar material.
The earliest published spectra showed that the spur consists of predominantly redshifted material.
The RFP scans presented in this paper confirm this result and show that the spur
traces a semi-coherent structure at redshifted velocities, moving progressively inward
(toward the center of \g292) with increasing radial velocity.  This trend is consistent with the geometric projection 
expected from an expanding shell.  A new feature, unseen in the earlier photographic images of \g292\,
(Goss \etal\, 1979; Tuohy, Clark \& Burton 1982) but clearly detected both in the [O~III] image (Figure~\ref{g292_oiii+allaps})
and our RFP scans (Figure~\ref{g292redblue}), is the collection of `streamers' seen radiating southward from 
the innermost edge of the spur.  The streamer emission also moves
inward with progressingly larger redshift, suggesting that these features trace a partially complete shell  
similar to the spur.

(2) An equatorial belt running E-W along the
projected center of the X-ray remnant.  The optical belt appears as a clumpy
line of broken filaments in Figure~1.  It is kinematically resolved
into multiple components by our RFP observations in Figure~2: a bar-shaped structure
of material at zero radial velocity, projected onto a high velocity component of clumpy
material.  Due to its intrinsic faintness, the belt emission escaped detection in the earlier
photographic plate images of \g292\, (Goss \& Shaver 1979; Tuohy,
Clark \& Burton 1982).  As we will show, the bar seen in our RFP scans appears
to be the optical counterpart to the bar seen in the \chan\, images by
Park \etal\, (2002, 2004) and Gonzalez and Safi-Harb (2003).
The appearance of both the bar and the diffuse
emission over most of the field at zero radial velocity suggests
that the bar also lies in the \g292\, rest frame.  This is
consistent with a slow wind (V\,$\lesssim\,$10 \kms) origin for the
bar.

(3) Fast-moving knots (FMKs, in analogy to similar features in Cas
A; Chevalier \& Kirshner 1978) seen in [O~III] (but not H$\alpha$) out
to radial velocities $\sim\,\pm\,$1500 \kms.  These conspicuous
features are clearly seen in the narrowband [O~III] image
(Figure~\ref{g292_oiii+allaps}), above and below the equatorial belt.
In the RFP data the FMKs range in size from around 5\arcsec\, down to
the seeing limit of 3\arcsec.  A clear trend in the FMK velocity
distribution of Figure~\ref{g292redblue}\, is that nearly all of the
northern FMKs are either seen at zero velocity or are blueshifted,
while most (but not all) of the southern FMKs are either seen at zero
velocity or are redshifted.  Unlike what is observed in Cas A (Reed
\etal\, 1995; Lawrence \etal\, 1995; Fesen \etal\,2001), the FMKs in
\g292 are more isolated and sparsely distributed across the remnant.
Aside from the spur, there are no networks of curved ejecta filaments
associated with the FMKs as seen in Cas A (Fesen \etal\, 2001).

\begin{figure}
\epsscale{0.6}
\plotone{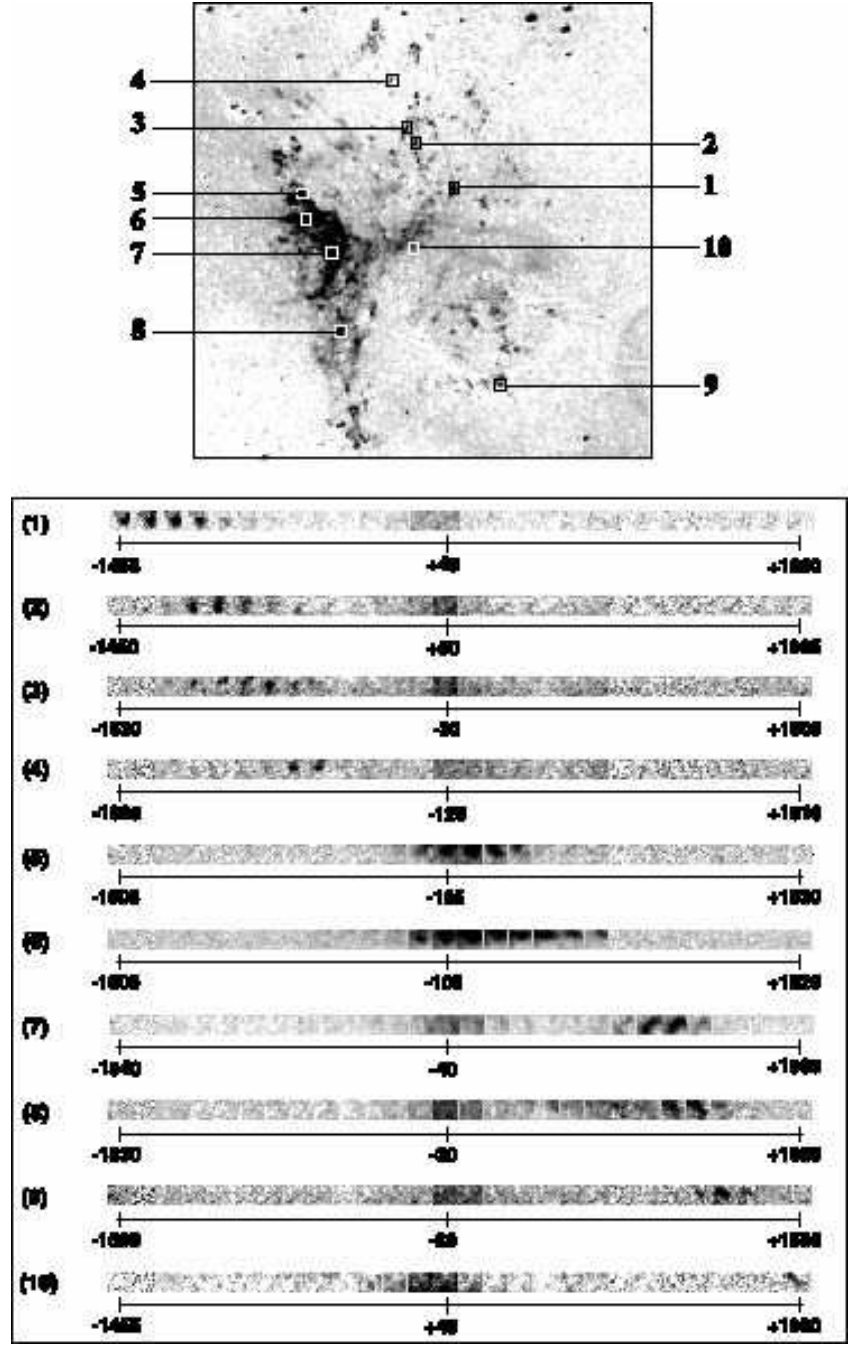}
\caption{The bottom panel shows a sequence of ejecta knot images from
the RFP data in the light of [O~III] arranged side-by-side in
velocity. The extraction regions are marked on the narrowband image of
\g292\ from WL05 (top panel).  Each square window in the bottom panel
is 11\arcsec\,$\times$\,11\arcsec\ in size and the individual frames
are separated by approximately 120 \kms\,
in velocity.  Radial velocities are marked at the bottom of each
frame.  A full resolution
figure can be found at http://fuse.pha.jhu.edu/\~{}parviz/papers/g292 .
}
\label{g292.2dknots}
\epsscale{1.}
\end{figure}

Although the velocity image sequence is a useful tool for exploring
the ejecta emission on a global scale, exploring the kinematics of
individual knots and the dynamical processes affecting them (e.g.,
shear, dynamical instabilities, etc.) is easiest when a sequence of
ejecta knot images are arranged side by side in velocity as
2-dimensional spectra.  We used the continuum-subtracted RFP scans to
construct an image sequence of selected features including FMKs,
sections of the spur and portions of the equatorial belt.  In
Figure \ref{g292.2dknots}\, we present a set of ten spectra consisting
of images 11\arcsec\,$\times$\,11\arcsec\, in size.  The sequence
includes the most blueshifted and redshifted knots detected in the RFP
images (K1 and K2, corresponding to spectra 1 and 10, respectively).
There is clear evidence of velocity shear seen in spectrum 6 from the
spur: emission extending redward from $-$100 \kms\, to +600 \kms.  In
the remaining spectra the faint, photoionized gas surrounding \g292\,
is clearly visible near zero radial velocity (particularly in spectra
1 and 9).  In spectrum 3 a pattern
can be seen in the FMK emission where the top of the clump first
appears at around $-$1200 \kms, then the lower portions progressively
appear at increasingly redshifted velocities.  Spectrum 10 illustrates
the spatial superposition between the zero velocity bar emission
and the high velocity ejecta (knot K1) present in the equatorial belt.

\section{SPECTRA AND [O~III] LINE PROFILE FITS}

To perform a detailed kinematic analysis of \g292\, we extracted spectra from both the spur
and the individual FMKs and fit the [O~III] line profiles.
Although the star subtraction process used to create the single velocity
images in Figure \ref{g292redblue}\, was adequate for the task of revealing the ejecta distribution,
we utilized the unsubtracted frames for the spectral extraction to better calculate the local background
around each emission knot.  We selected FMKs for spectral extraction by eye, taking care to avoid
stellar contamination by checking that knots in the RFP scans did not appear as
stars in the DSS blue image of the \g292\, field. 
The apertures used for the spectral extractions were 3\farcs2$\times$3\farcs2 squares, roughly equal to the seeing
FWHM of the smoothed data.  The extraction apertures for the bright southeastern spur were of the same
size, laid out on a grid covering the entire structure.  The extraction apertures also covered the brightest
knots of streamer material seen just south of the spur in Figure~\ref{g292_oiii+allaps}.

The spectral extraction consisted of first summing the emission within each defined extraction aperture,
for each frame of the datacube.  
To subtract the sky emission, we selected a ring of pixels in each frame, centered
on the optical axis and passing through the middle of each extraction aperture.  This ensured that
the sky fluxes were extracted at the same wavelength as the emission within each
aperture.  The ring consisted of all the sky lying off the supernova remnant which included
the widespread diffuse emission but excluded stars or other detectable discrete sources.  We summed
the sky emission along the ring, rejecting pixels deviating more than 4$\sigma$ from the mean value.
We then multiplied the resulting sky spectrum by $N_{pix}(obj)/N_{pix}(sky)$
to obtain a scaled sky spectrum for each of the apertures.
Subtracting this spectrum from that of each emission knot yielded the final spectrum for each
aperture.  The sky subtraction removes large scale sources of background emission in the
RFP data bandpass, namely Galactic [O~III] and diffuse [O~III] surrounding \g292.  

Routines were written to extract the [O~III] profiles and subtract background emission from 
all the spectra.   Compared to typical long-slit spectral observations (e.g., the Cas A study by Reed 
\etal\,1995), the spectral range of our data is much shorter, covering only $\sim\,$52\AA\, at [OIII] and 
$\sim\,$30\AA\, at H$\alpha$. In addition, our spectra are typically less than critically sampled.  We
fit our data with a spectral profile that included both the instrumental and expected intrinsic source
profile shapes, as described below.  Due to the large number of 
spectra obtained from both the FMKs and the spur region, it was necessary to automate the line identification
and fitting process.

Most of the FMK spectra  exhibited solitary [O~III] emission components.  However,
due to the combined factors of geometrical projection and the possible action of dynamical processes within
the shocked ejecta of the spur (instabilities, shear motions, etc.),
the RFP spectra of the spur often exhibited [O~III] emission from multiple velocity components.
We did not know {\em a priori} the number of [O~III] lines to expect in any single RFP spectrum,
and therefore needed a method of determining the number of distinct significant emission lines in each spectrum.
Initial trial and error showed that maxima
less than 4$\sigma$ above the background failed to yield stable line profile fits, so only
features with higher S/N were counted as separate components.  
Although this eliminates the possibility of measuring
radial velocities for some of the fainter emission, we were more concerned here with making reliable
line identifications and measuring accurate radial velocities for well-defined features.
After rejecting spectra of some of the fainter features in the RFP data, our final sample consisted 
of 62 FMK and 462 spur/streamer spectra.   We assumed the [O~III] profiles
to be intrinsically Gaussian in shape; convolved with the instrumental response of the RFP the
resulting profile is a Voigt function.  Therefore, we fit emission lines
with Voigt functions of fixed Lorentzian width (the instrumental value, 52 \kms) and left the line 
flux, Gaussian width, line centroid and background level as free parameters.  During each profile fit
we initially set the width of the Gaussian component equal to the instrumental value and the line centroid
equal to the velocity of the local flux peak.  

\begin{figure}
\epsscale{0.7}
\plotone{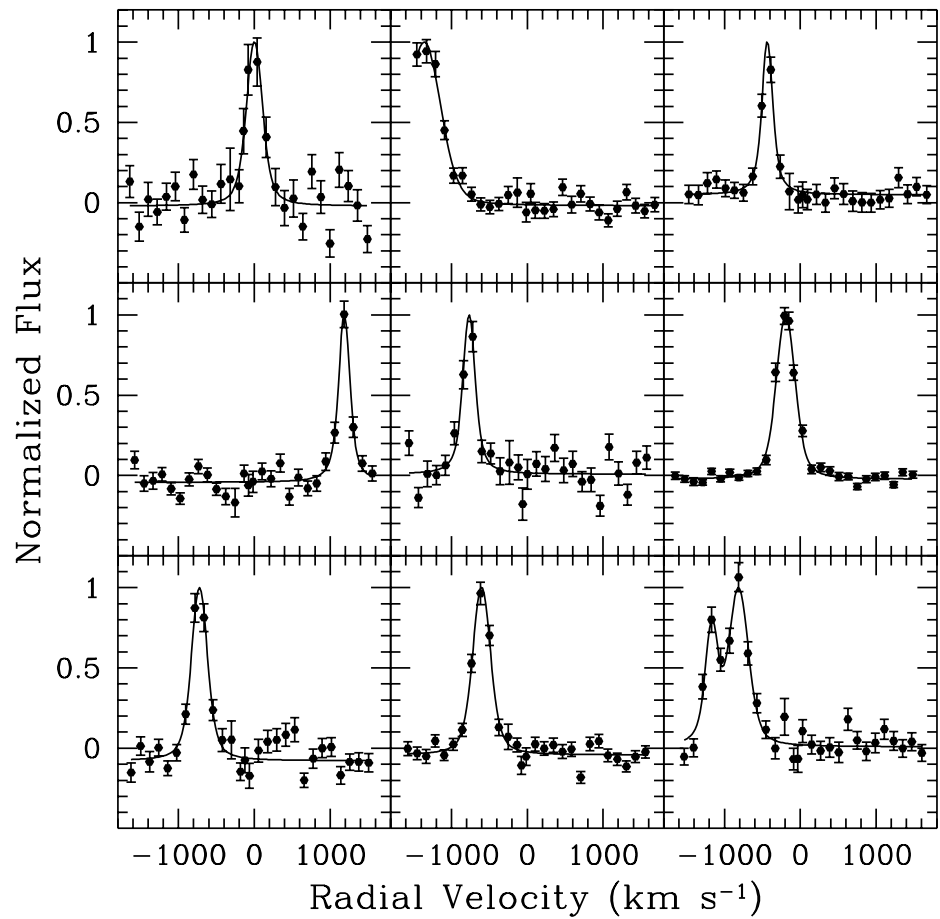}
\caption{The [O~III] profiles of 9 selected FMKs in \g292.  The solid
line drawn through the data is the best fit to the line profile.  Line
profiles are shown over the full range of velocity shifts and include
both low and high S/N examples.  }
\label{g292knotspec}
\end{figure}

\begin{figure}
\epsscale{0.7}
\plotone{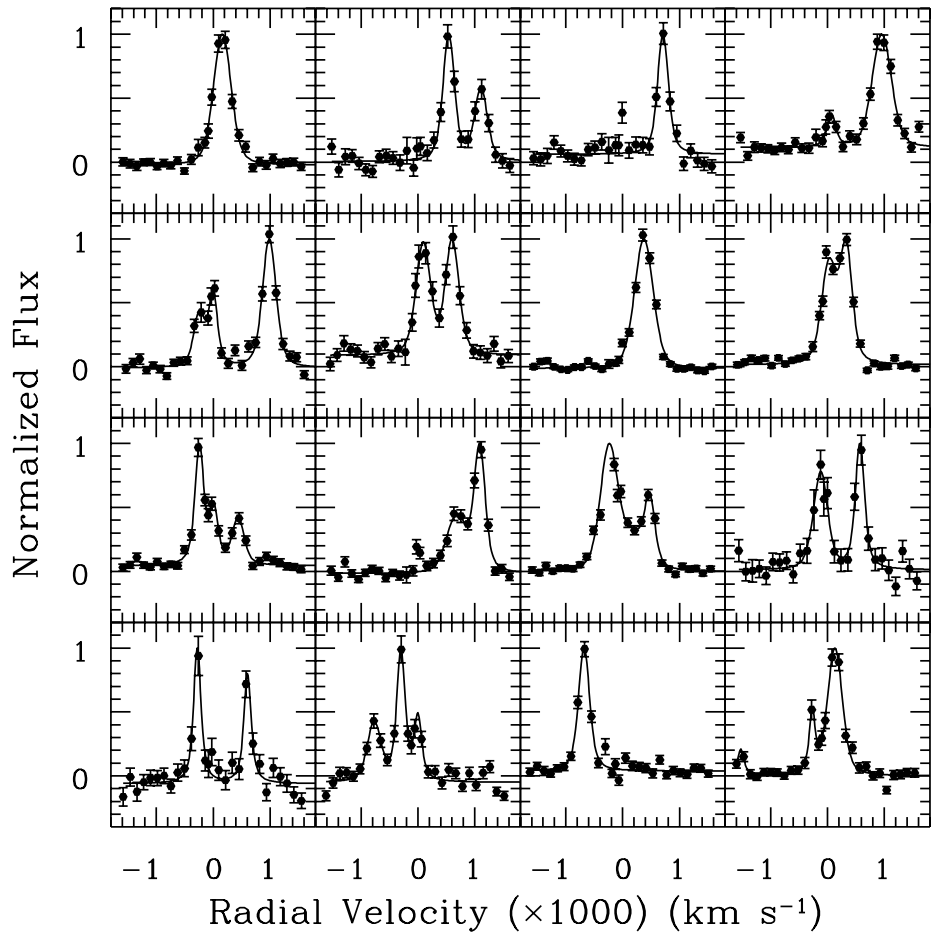}
\caption{The [O~III] profiles of 16 selected locations in the bright
eastern spur region.  The solid line drawn through the data is the
best fit to the line profile. }
\label{g292spurspec}
\end{figure}

The Gaussian widths yielded by our fits to all the [O~III] profiles
ranged from close to the instrumental Gaussian width (44 \kms) to
values around 100 \kms, with a few profiles exhibiting widths as large
as 350 \kms.  Overall, the line widths are
consistent with [O~III] emission from radiative shocks with
\vs\,$\sim$\,50$-$200 \kms, although the large widths of some of the emission lines may be
produced by blending of narrower features.  In Figures \ref{g292knotspec}\, and
\ref{g292spurspec}\, we present examples of spectra from nine FMKs and
sixteen locations in the spur, along with the line profile fits
obtained using the method described above.

Although the main goal of the RFP observations was to map the [O~III]
$\lambda$5007 emission in \g292, it is not known whether other
emission lines could have been shifted into the bandpass of our
spectra.  Optical studies of other OSNRs such as Cas A (Reed \etal\,
1995) have shown that the emission lines from some ejecta knots may be
Doppler shifted enough to cause ambiguity in distinguishing lines of
one atomic species from those of another.  Obviously this is also a
potential concern for \g292, where some features in our RFP spectra
may be classified as either highly blueshifted [O~III] $\lambda$5007
emission or highly redshifted [O~III] $\lambda$4959 emission.  In this study
we assume that all the bright spectral lines we have fitted are
due to the [O III] $\lambda$5007 transition. It is possible
that this has produced an error in the computed radial
velocity (by $+$2900 km s$^{-1}$ if we have misidentified
as $\lambda$5007 what is actually a $\lambda$4959 transition).
Resolving this potential ambiguity would require additional spectroscopy
over a wider spectral band. For the ensemble of knots, however, we believe that
this is unlikely to be an important or widespread effect.
As we show in the next section, the identification of the FMK spectral
features as $\lambda$5007 lines leads to a simple
kinematic picture of the FMKs, and yields an age for \g292\, consistent
with the age predicted by recent global models of the supernova remnant/PWN
energetics (Chevalier 2005).

\section{RADIAL VELOCITIES OF THE EJECTA DISTRIBUTION}

\subsection{FAST-MOVING KNOTS}\label{FMKs}

\begin{figure}
\epsscale{0.5}
\plotone{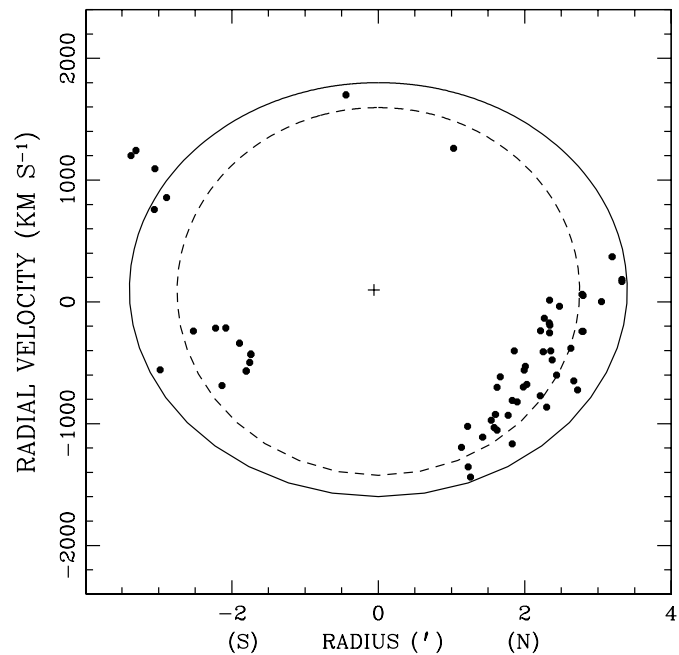}
\caption{ Radial velocity vs.~projected radius for fast-moving knots in \g292.
Radii were measured from the geometric center of the radio image of
the remnant (GW03).  The solid line marks the best
fit velocity ellipsoid that encompasses the outermost points of the
velocity distribution, excluding the most redshifted southern
FMKs. The dashed line marks the best fit averaged through all points
of the velocity distribution.}
\label{g292knotvelcircle}
\end{figure}

Using the [O~III] line profile fits, we calculated the radial
velocities of the 62 selected FMKs and plotted the velocities as a
function of the knots' radial distance from the geometric center
($\alpha$,$\delta$) = 11$^{\rm h}$24$^{\rm m}$34\fs 68,
$-$59$^{\circ}$15$^{\arcmin}$29$^{\arcsec}$ (2000) for \g292\, as
determined from the radio observations of GW03.
We present a position-velocity diagram of the FMKs in
Figure \ref{g292knotvelcircle}.  The correlation observed between FMK
velocities and their distances from the geometric center in
Figure \ref{g292redblue}\, can be clearly seen in
Figure \ref{g292knotvelcircle}.  The velocities of the northern FMKs
increase with radial distance, ranging from around 0 \kms\ at a 
1\arcmin\ radius to around 1500 \kms\ at the outer edge (3\farcm5) of the
sampled FMK distribution.  The northern FMKs appear to trace a portion of an ellipse
in the position-velocity plane, consistent with the kinematic
signature of an expanding shell.  Another identifiable property of the
velocity distribution is that nearly all of the northern FMKs exhibit
blueshifted emission, while, by contrast, the southern FMKs exhibit a
more even mixture of blueshifted and redshifted emission.

It is also clear from Figure \ref{g292knotvelcircle}\, that there is a
significant spread in FMK radial velocities ($\sim$500 \kms) at a
given radius.  This dispersion is much larger than the random
statistical error in velocity ($\sim$50 \kms).  The FMKs may then
not strictly lie on a geometrically 
thin shell, making the definition of a shell expansion radius less well defined.
However, we can at least fit a velocity ellipsoid to the outermost
FMKs in the position-velocity diagram in order to obtain a rough
expansion age for \g292.

We fit the position-velocity distribution in
Figure~\ref{g292knotvelcircle}\, with a velocity ellipsoid, leaving the
radius $R_{ej}$, expansion velocity $V_{ej}$ and systemic velocity
$V_{c}$ as free parameters.  The expansion center coordinates were
fixed to those of the geometric radio center of GW03.
Excluding the southernmost FMKs, we fit a curve to match
the outermost points of the radial velocity distribution as closely as
possible, paying particular attention to matching the most
well-defined pattern, that of the blueshifted northern FMKs.  Our best
estimate is marked by the solid curve in Figure \ref{g292knotvelcircle}.

The fitted radius of the FMK position-velocity distribution is
3\farcm4 (5.9\,$d_{\rm 6}$ pc).  This defines a circle with
northern edge lying just interior to the northernmost FMKs seen in
Figure \ref{g292_oiii+allaps}.  We find an FMK shell velocity
V$_{ej}\,=\,$1700 \kms, with a velocity centroid $V_{c}\,=\,$+100
\kms, although the latter is not very tightly constrained since it is
comparable to the random statistical error in FMK radial velocities
($\pm$50 \kms) and is smaller than the FMK velocity dispersion.
Our derived systemic shift is substantially smaller than the shifts seen
in the other known OSNRs: Cas A (+900 \kms; Reed \etal\, 1995),
N132D ($-$500 \kms; Morse \etal\, 1995), E0102$-$72.3
($-$500 \kms; Tuohy \& Dopita 1983), 0540$-$69.3
(+370 \kms; Kirshner \etal\, 1989) and the remnant in NGC 4449
(+500 \kms; Balick \& Heckman 1978).  
Assuming free expansion of the shell, the age $\tau$
($\equiv\,$R$_{ej}$/V$_{ej}$) is $3400\,d_{\rm 6}$ years.

As a separate exercise, we also computed $\tau$ 
for two separate cases: (1) assuming the FMKs formed with
a range of initial velocities but experienced negligible deceleration over
the lifetime of the SNR, and (2) assuming the FMKs formed with identical
velocities but with different knots experiencing differing degrees of
deceleration over the lifetime of the SNR.  In case 1 we fit the position-velocity
distribution of all the FMKs, including the knots left out of the previous fit.  In
that case, we found a radius of 2\farcm75 ($4.8\,d_{\rm 6}$ pc),
a shell velocity of 1500 \kms\, and $V_{c}\,=\,$+85 \kms\, (this fit is
marked by the dashed ellipse in Figure \ref{g292knotvelcircle}).  These
numbers give $\tau=3100\,d_{\rm 6}$ years.  In case 2 we did not know
the degree of deceleration of each individual knot, so we took the 500 \kms\
spread in the FMK velocity distribution as an approximation of the maximum
deceleration experienced by the outermost knots.  In that case, the
original speed of the outermost knots was $V_{0}$\,=\,$V_{ej}\,+\,\Delta\,V$\,=\,
1700\,+\,500 \,=\,2200 \kms, and the expression for the age is now
$\tau\,\equiv\,$R$_{ej}$/(V$_{0}\,-\,\frac{1}{2}\,\Delta\,$V). Taking 
$R_{ej}$\,=\,3\farcm4 (5.9\,$d_{\rm 6}$ pc), the resulting age is $\tau=3000\,d_{\rm 6}$ years.
Thus, the range in ages derived from our analysis is (3000-3400)\,$d_{\rm 6}$
years.

The kinematic age derived from our analysis is nearly twice the value
inferred by Murdin \& Clark (1979) from spectroscopy of the spur.  The
disparity between our age and that of Murdin \& Clark (1979) occurs
because the expansion radius of the FMKs (3\farcm4) is nearly 70\%
larger than that of the spur ($\sim$2\arcmin), while the expansion
velocities of the two distributions are nearly the same
($\sim\,\pm$\,1700 \kms).  As we illustrate in the next section, the
position-velocity diagram of the eastern spur is highly complex and
structured.  It is unclear whether a simple geometric expansion model
can be applied to the spur; therefore, we believe that modeling the
FMK kinematics provides cleaner, better estimates of both the
expansion velocity and age of \g292.

Recently Chevalier (2005) has applied self-similar models to constrain
the supernova explosion types of a number of known core collapse SNRs,
including \g292.  The models use observationally determined parameters
such as the pulsar period and spindown rate, PWN luminosity and blast
wave radius to classify the natal explosion type of each SNR.  In the
case of \g292\ assuming a distance of 6 kpc, Chevalier (2005) found
that an age of 2700--3700 years was required for consistency with his
models.  Another estimate of the remnant's age, 2900 yrs, comes from
the spindown age of the pulsar, PSR~J1124$-$5916, associated with
\g292\ (Camilo \etal~2002).  These values are in good agreement with
the age of (3000--3400)$\,d_{\rm 6}$ years from our fit to the FMK
radial velocities in \g292. This consistency in age estimates supports
our assumption that the outermost knots in the position-velocity
distribution have not been strongly decelerated.

\subsection{THE SPUR}

\begin{figure}
\epsscale{0.5}
\plotone{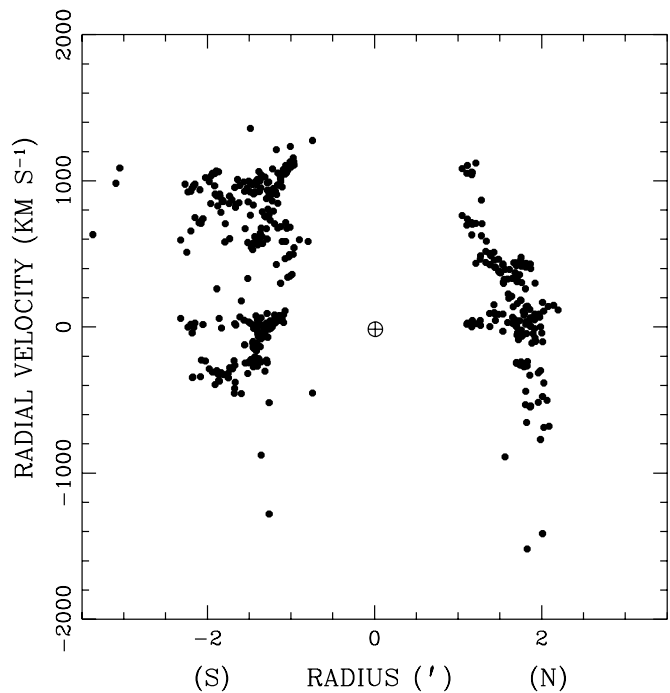}
\caption{Radial velocity vs.~projected radius for 462 points in the spur
region of \g292. The radii were determined as in Figure~\ref{g292knotvelcircle}.}
\label{g292spurvelcircle}
\end{figure}

The position-velocity diagram of the spur (Figure~\ref{g292spurvelcircle})
presents a far more complex kinematic picture than that of the FMKs
(Figure~\ref{g292knotvelcircle}).  Although the spur does exhibit some of the overall
properties expected from an expanding shell (e.g., the inward
progression of the [O~III] emission with increasing radial velocity seen in
Figure \ref{g292redblue}), one is hard pressed to identify the kinematic
signature of such a shell in Figure \ref{g292spurvelcircle}.  
One place where such a kinematic signature possibly appears is in
the redshifted ejecta north of the expansion center.  
Between radii of 1\arcmin\, and 2\arcmin, the velocity of this material
monotonically decreases from +1000 \kms\, to 0 \kms.  A more prominent
feature is the tendency for many of the points in the
position-velocity diagram to aggregate into radially extended 
clusters at nearly constant velocity.  In particular, as demonstrated in
Figure \ref{g292spurvelcircle}, there are two clusters of points (mostly
toward the south) located
at roughly 0 \kms\, and +950 \kms\, (both noted by Braun \etal\,
(1983)\footnote{Braun \etal\, (1983) employs the word `spur' to describe only
the radially extended, constant velocity features traced in the
optical spectra, whereas we use the word as a descriptive term for the
morphology of the O-rich material.}), with a similar but less distinct
clustering possibly at $-$300 \kms.  Braun \etal\, (1983) interpreted these
features as 'breakouts' of O-rich clumps superimposed on a more
uniform, expanding shell of ejecta.  The RFP data
appear to be consistent with this interpretation.
Despite the rather confused radial velocity
picture of the spur, both this feature and the FMK distribution
described earlier extend over a total velocity range of approximately 3000
\kms.

PSR J1124$-$5916 in \g292\ is intriguingly close to the inner edge of the O-rich spur 
(marked in Figure \ref{g292_oiii+allaps}).  It is offset $\sim$45\arcsec\, eastward
from the geometric center of the SNR, implying a substantial velocity
component (V$_{\perp}\,\sim$ 450 \kms\, for $\tau\,=\,$3400 years) in the
direction of the spur.  This feature,
combined with the known presence of a luminous PWN around PSR
J1124$-$5916 raises an interesting question about the origin of the
spur emission: could the spur be a shell of dense ejecta swept up from
the inside by the motion of PSR J1124$-$5916?

There have been no unambiguous kinematic signatures
predicted for the interaction between a PWN and SN ejecta.  However,
vital clues may be found in existing optical spectra of this OSNR: 
we would expect that as PSR J1124$-$5916 
moved through the interior of \g292\, the innermost ejecta (mostly Ar-Fe
produced in explosive nucleosynthesis) would be the first material to be shocked
by the PWN.  The lack of line emission from elements heavier than S in the global
optical spectra of \g292\, (Murdin \& Clark 1979; Sutherland \&
Dopita 1995; WL05) strongly indicates that the optical emission from the spur
does not arise in a PWN/O-rich ejecta interaction.   As mentioned earlier, X-ray
emission from these heavier species is also absent in the \chan\, spectra of
the \g292\, ejecta (Park \etal\, 2004), suggesting that in general there
has been little or no interaction between the PWN and SN ejecta.  A strong
contrast may be drawn between here and the case of the OSNR 0540$-$69.3 in
the LMC, where the PWN observed in the X-rays is completely enclosed by a ring of [O~III]-emitting
ejecta (Reynolds 1985).  There, optical emission from S, Ar, Fe and Ni are observed
in addition to the usual O and Ne lines (Kirshner \etal\, 1989; Serafimovich
\etal\, 2004).  Based on evolutionary considerations, Reynolds (1985) ascribed
the optical emission from 0540$-$69.3 to a PWN/ejecta interaction, consistent
with the presence of optical emission from the innermost ejecta.

\begin{figure}
\epsscale{.7}
\plotone{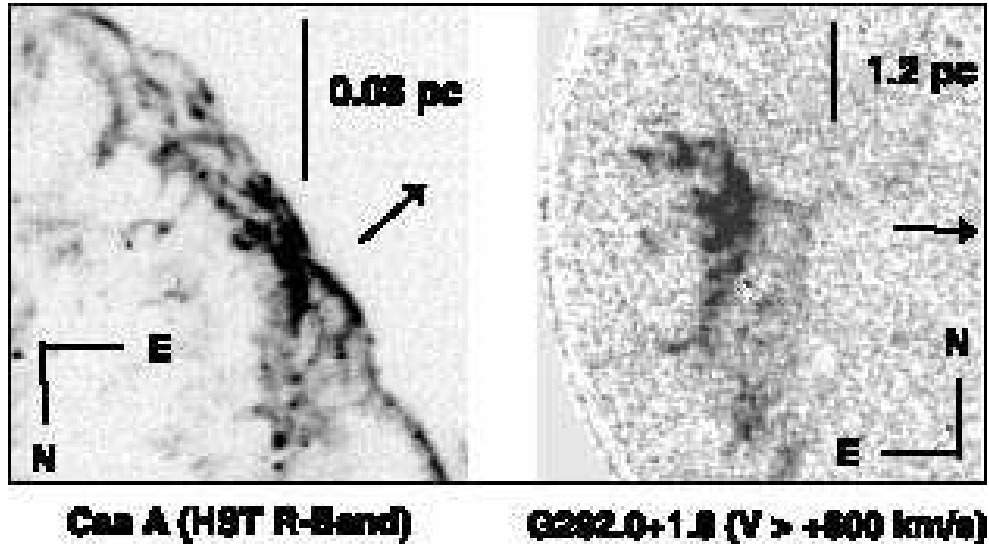}
\caption{HST image of shocked ejecta in the northwest portion of Cas A
(left panel, from Fesen \etal~2001) displayed beside an image of
\g292\ (right panel) showing the summed [O~III] emission from all
redshifted RFP scans (V\,$\geq\,$+600 \kms).  For ease in comparing
morphologies of the ejecta shells, the Cas A image has been rotated so
that the expansion centers of the two SNRs (marked by the arrows) lie
in roughly the same direction.  Scale bars are shown for assumed
distances of 3.4 kpc (Reed \etal~1995) for Cas A and 6 kpc (GW03)
for \g292.  Despite the order of magnitude difference
in spatial scale, both SNRs exhibit scalloped, concave filaments
pointing away from the explosion center as would be expected for
Rayleigh-Taylor unstable material recently compressed by the reverse
shock.  A full resolution
figure can be found at http://fuse.pha.jhu.edu/\~{}parviz/papers/g292 .
}
\label{g292casA}
\end{figure}

The remaining interpretation for the spur emission in \g292\, is that 
it arises in an overdense shell of ejecta overtaken by the reverse shock. Evidence
for this picture may be found by comparing the morphology of
the spur with HST optical images of Cas A (Fesen \etal\, 2001), the
most well-studied OSNR.  Although the size scales of the Cas A and
\g292\, filaments differ by over an order of magnitude, they bear a
striking similarity: both follow a concave shape, curving away from
the SNR expansion center, and both exhibit the scalloped morphology
characteristic of Rayleigh-Taylor instabilities (Figure
\ref{g292casA}).  Most importantly, the `wiggles' in the spur point
away from the expansion center, in close analogy to similar features
seen in Cas A.  This suggests that high density material is
penetrating into overpressurized, less dense material, precisely the
condition required to produce Rayleigh-Taylor instabilities. 
These instabilities would likely leave a significant kinematic imprint
on the ejecta velocity distribution, and may be partly responsible for
much of the dispersion seen in the position-velocity diagram of the spur
(Figure \ref{g292spurvelcircle}).

\section{LOW VELOCITY OPTICAL EMISSION}

\subsection{H$\alpha$ EMISSION}

\begin{figure}
\epsscale{0.7}
\plotone{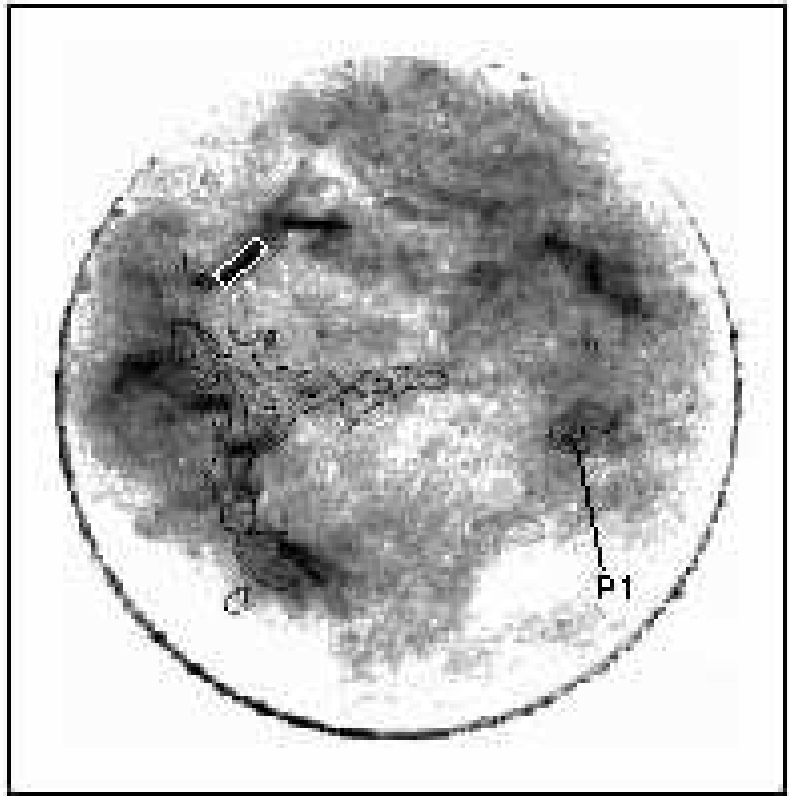}
\caption{The zero-velocity H$\alpha$ image shown with logarithmic
contours from the zero velocity [O~III] image in
Figure \ref{g292redblue}. There is little direct correspondence between
[O~III] and H$\alpha$ emission from the eastern spur, confirming that
this feature is ejecta material. The extraction box for the spectrum
in Figure \ref{g292halspec}\, is marked on the filament with the white
box. A full resolution
figure can be found at http://fuse.pha.jhu.edu/\~{}parviz/papers/g292 .
}
\label{g292haloiii}
\end{figure}

After following the procedure described earlier for subtracting stars
from the RFP scans, we examined the H$\alpha$ datacube for evidence of
hydrogen and nitrogen line emission from \g292.  The only scans
exhibiting structure were the zero velocity scan and its two adjacent
frames centered on $-$62 \kms\, and +85 \kms.  Interpolating the flux between
RFP frames, we produced an image of the entire field at zero radial
velocity.  In Figure \ref{g292haloiii}\, the continuum-subtracted zero
velocity H$\alpha$ image is shown along with contours from the
zero-velocity [O~III] image.  Overall there is very little detailed
correspondence between the H$\alpha$ and [O~III] morphologies of
\g292.  In particular, the high-velocity [O~III] structures observed 
at all velocities in the RFP images exhibit little or no corresponding
H$\alpha$ emission.

The most prominent features in the H$\alpha$ image are a set
of narrow filaments 30\arcsec\,$-$40\arcsec\, in length
seen mostly on the eastern half of the H$\alpha$ image.
The filaments lie close to the O-rich spur, but due to the
grossly dissimilar morphologies of the H$\alpha$ and [O~III]
emissions, it is unlikely that they trace the same material.  The
filaments appear at zero radial velocity (i.e., in the rest frame of
\g292), which raises the possibility that they are clouds of either
shock-excited or photoionized circumstellar material.  The lack of
obvious emission from these filaments in both the [O~III] RFP scans
and the \chan\, image of \g292\, places a strong constraint on the
excitation mechanism of the H$\alpha$ filaments: we can rule out both
radiative shocks in O-rich material and radiative shocks faster than
100 \kms\, in ISM material (Hartigan, Raymond \& Hartmann 1987).

\begin{figure}
\epsscale{0.5}
\plotone{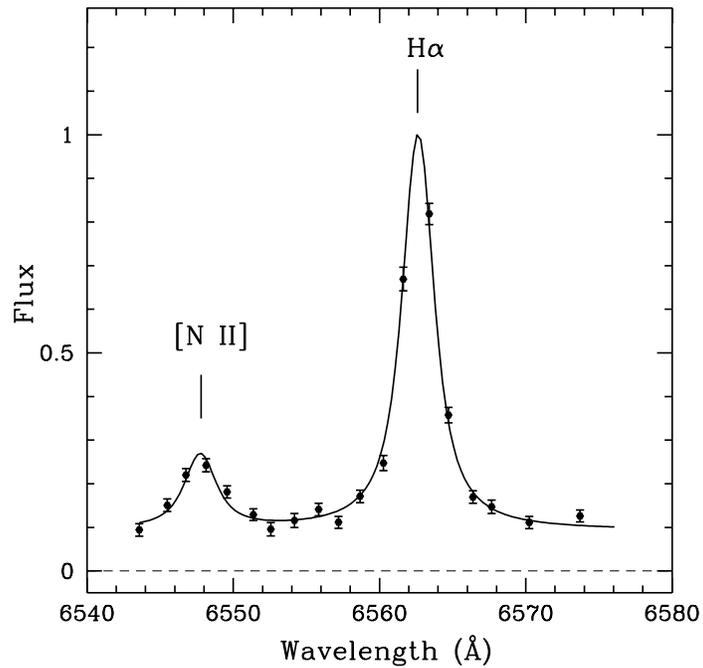}
\caption{Sky-subtracted spectrum of a filament in the zero velocity
H$\alpha$ image. Residual continuum is due to the presence of stars
along the filament.  The solid curve shows the best fits to the
H$\alpha$ and [N~II] $\lambda$6548 line profiles.  }
\label{g292halspec}
\end{figure}

If the filament emission is produced in slow ($<$ 100 \kms) radiative shocks,
then we should be able to match the [N~II]:H$\alpha$ flux ratios with the
theoretically predicted values.  We obtained a spectrum from one of the filaments
(the extraction box is marked in Figure~\ref{g292haloiii} and the spectrum is plotted in
Figure~\ref{g292halspec}) and fit the H$\alpha$ and [N~II] $\lambda$6548
line profiles.  Both lines were fit by Voigt profiles with widths
fixed at the instrumental values.  To within the errors ($\pm$15 \kms)
the H$\alpha$ and [N~II] $\lambda$6548 lines are centered on zero
radial velocity, consistent with an origin in the circumstellar medium
around \g292.

From the profile fits we find that the ratio
[N~II]($\lambda$6548):H$\alpha$\,$\approx$\, 0.2, with spectra
extracted from the other nearby filaments exhibiting nearly the same
ratios.  Theoretical values of the [N~II]($\lambda$6548):H$\alpha$
ratio have been calculated by Hartigan, Raymond \& Hartmann (1987) for
plane parallel shocks with equilibrium preionization and cosmic
abundances.  The ratios range from approximately 0.1 for a 90 \kms\,
shock (higher values can be rejected because [O~III] emission turns on
very rapidly for faster shock speeds) down to nearly zero for a shock
speed of 20 \kms.  For this scenario to be correct, we therefore
require significant enhancement (by factors of 2 or more) in the
nitrogen abundance in the ambient medium.  This explanation also 
requires very high densities in the emitting filaments.  The shock
parameters of the \g292\, blast wave ($n_{0}\,\sim\,$1 cm$^{-3}$,
\vs\,$\sim$\,2000 \kms; GW03) imply a high ram pressure,
which combined with pressure conservation and low shock speed in the
filaments ($\lesssim$90 \kms) results in predicted preshock cloud densities $\gtrsim$100
cm$^{-3}$.  The presence of such dense, localized structures in a
predominantly low density medium seems unlikely, but it may 
be possible if the progenitor wind was highly clumped.

An alternate explanation for the filaments is that they
are low density (n$\,\sim\,$1 cm$^{-3}$), neutral clouds photoionized
by radiation from the shocks in \g292.  As noted by Hamilton \& Fesen
(1986) for SN 1006 and Morse \etal\, (1996) for N132D, the shocked 
metal-rich ejecta inside a supernova
remnant is a source of hard UV and X-ray line radiation which escapes
out into the ISM and ionizes material surrounding the SNR.  Another
source of hard radiation is He~II $\lambda$304 photons from the ISM
material swept up by the blast wave.  Exposed to the dilute, time
varying radiation field of the SNR, nearby H~I clouds could remain
significantly neutral while being heated to temperatures
$\sim$12,000$-$15,000~K (Ghavamian \etal\, 2000).  The resulting
optical emission from the cloud would feature faint H$\alpha$, [N~II],
[S~II] and H$\beta$ emission and almost negligible [O~III] (as seen in
Tycho's SNR, Ghavamian \etal\, 2000).  Unlike the case of radiative
shocks, the photoionization scenario can produce the observed
[N~II]:H$\alpha$ ratios without the need for enhancing the nitrogen
abundances in the emitting clouds.  A long-slit spectrum with larger
wavelength coverage and higher spectral resolution is required
to definitively establish the physical properties of the filaments in
Figure \ref{g292halspec}\, and to decide between the photoionization and
radiative shock scenarios.

As we have already mentioned, there is an absence of high-velocity
H$\alpha$ and [N~II] emission over the dynamical range sampled by our
RFP data.  This sort of material has been seen in Cas A (the
`fast-moving flocculi', or FMFs, reported in Fesen, Becker \& Blair
(1987)).  The Cas A FMFs exhibit properties similar to both
FMKs and QSFs, namely high velocities and emission from
nitrogen, hydrogen and helium.  They are believed to be fragments of the
progenitor photosphere which were expelled at high velocity during the
supernova explosion.  Since the photosphere comprises the outermost
layer of the ejecta, it is the first material to be overrun by the
reverse shock. \g292\, is nearly ten times older than Cas
A, while the survival time of dense, shocked knots is typically only on the order
of a few decades. It seem likely then that any hydrogen/nitrogen-rich material ejected from
the surface of the progenitor star would have been overrun and destroyed at an earlier
evolutionary phase of \g292.

\subsection{OPTICAL EMISSION FROM THE EQUATORIAL BAR}

In our optical investigation of \g292\, it is perhaps our study of the
equatorial belt that benefits the most from the kinematic resolving
power of the RFP.  As previously defined (see \S\ref{global}), the
equatorial belt appears in the zero radial velocity [O~III] image as a
narrow bar of emission elongated in the E-W
direction. This is very different from the larger, clumpier structure
seen in the narrowband image in Figure \ref{g292_oiii+allaps} due to
the geometric projection of high velocity knots (such as K1 and K2 in
Figure \ref{g292_oiii+allaps}) onto the zero-velocity bar.

\begin{figure}
\epsscale{0.7}
\plotone{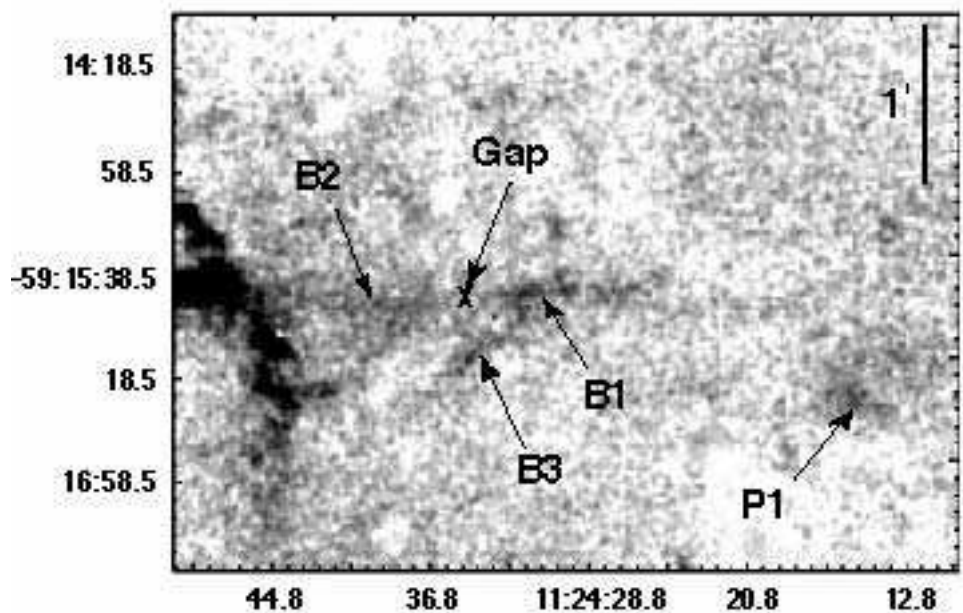}
\caption{Closeup of the equatorial bar region in \g292~in zero radial
velocity [O~III] from the RFP scans (bottom).  The features discussed in the
text are marked.  The 'X' marks the position of the geometric radio center 
measured by GW03.  A full resolution
figure can be found at http://fuse.pha.jhu.edu/\~{}parviz/papers/g292 .
}
\label{g292_markbelt}
\end{figure}

We can trace the detailed structure of the belt at zero velocity using
the marked features in Figure \ref{g292_markbelt}.  In this image the
most prominent portion of the bar (B1) lies immediately west of the radio
geometric center and is approximately 5\arcsec\, thick and
22\arcsec\, long.  The [O~III] surface brightness of B1 is $\sim$20\% of the brightest
emission in the [O~III] spur.  Fainter bar emission can also be seen
extending to the east (B2) and southeast (B3) of B1. Both the bar
and a patchy feature 2\arcmin\, further southwest (marked as P1
in the figure) are visible in our H$\alpha$ image, although the 
bar is relatively fainter and smoother in H$\alpha$ than in [O~III].
As we will show in Section 8.2,
features B1 and B2 appear to be optical counterparts to portions of the 
X-ray bar seen first in the \eins\, and then the \chan\,
image of \g292.  On the other hand, another prominent zero-velocity
feature seen in Figure \ref{g292_markbelt}, B3, does not exhibit a
clear X-ray counterpart.

\section{COMPARING THE X-RAY, OPTICAL, AND RADIO EMISSION FROM \g292 }

The RFP data of \g292 shows that the [O~III] emission is distributed
over a wide range in velocity and consists of multiple components.
Therefore, to understand the interaction of \g292\,
with its surrounding medium we must separate the ejecta knot [O~III] emission
from circumstellar/ambient [O~III] emission.
Since much of the emission from \g292\, is emitted in the X-ray band,
comparing our RFP data with the spectacular \chan\, ACIS image of
\g292\, provides a link between the optically-emitting components and global
variations in density, temperature, composition and ionization state. 
For this comparison we utilized a 38 ks
archival \chan\ observation of \g292, filtered to include only the
oxygen K line emission between 0.55 and 0.75 keV.  We then smoothed
the image with a boxcar of variable size, preserving a minimum S/N of
3 per smoothing beam.  The smoothed 0.55--0.75 keV image of \g292\,
is shown in the left panel of Figure \ref{g292opticalxray}. The X-ray
contours are also shown to provide a reference for the eye in
distinguishing contours around emission minima from contours around
emission maxima.  Contours of the 0.55$-$0.75 emission are overlaid
onto the zero velocity [O~III] and H$\alpha$ images in the middle and
right panels, respectively.

\subsection{TRACING THE EJECTA DISTRIBUTION}

An important question arising from our RFP analysis is how the
fast-moving ejecta seen in [O~III] correlate with the chaotic X-ray
emitting material seen in the \chan\, ACIS image of \g292.
The answer appears to depend on whether the optical emission in
question belongs to the FMK population, equatorial belt or eastern
spur material.  In the following discussion, we present a qualitative first
comparison between the optical and X-ray ejecta emission in \g292.

{\it FMKs $-$}\,There is a clear spatial correlation between some FMKs
seen in the [O~III] RFP data of \g292\, and the knots of X-ray ejecta
seen in the \chan\, image (Figure~\ref{g292fmks_xray+oiii}).  In
particular, at the positions of the northern (blueshifted) knots, the
extended X-ray features invisible at optical wavelengths often appear
to terminate in FMKs of [O III] emission lacking X-ray counterparts.
In addition, some of the [O~III] FMKs appear to be `nested' at
intersection points between elongated strands of X-ray-emitting
ejecta.

The correlations described above are consistent with ram
pressure conservation in ejecta fragments of variable density.  When
an ejecta fragment encounters the overpressured gas behind the reverse
shock, there are radiative shocks (V$_{S}$\,$\sim$100$-$200 \kms) driven
into the densest portions of the clump (n$\gtrsim\,$100 cm$^{-3}$),
while partially radiative shocks (V$_{S}$\,$\sim\,$500 \kms) form in
locations of intermediate density (n\,$\sim\,$10 cm$^{-3}$) and fast
non-radiative shocks (V$_{S}$\,$\gtrsim\,$1000 \kms) form in least
dense (n\,$\lesssim\,$1 cm$^{-3}$) portions of the clump.  In the
resulting mixture of conditions, the brightest optical and X-ray
features often follow mutually exclusive spatial distributions, while
material of intermediate surface brightness are often detected in both
bands.

Whereas in a number of northern FMKs we can trace what appear to be connected
strands of ejecta material between the optical and X-ray images, most of the
southern FMKs tend to appear as solitary localized structures in
[O~III] surrounded by more extended, diffuse X-ray emission
(Figure~\ref{g292fmks_xray+oiii}).  The dissimilarity in the
X-ray-optical correlation between the northern and southern FMKs may
reflect a comparatively larger density contrast in the southern ejecta
distribution: most of the southern knots may be composed of material
too dense to produce anything but radiative shocks.  It is unclear
whether this inferrence, if correct, points to variations in the
pattern of hydrodynamical instabilities during the SN event or whether
it reflects differences beween the history of the reverse shock/ejecta
interaction in the northern and southern portions of \g292.

A large density contrast between the optically emitting and X-ray
emitting material in the southern FMKs would have consequences on the
morphology and kinematics of the knots.  Using hydrodynamical
simulations, Klein, McKee \& Colella (1994) and Jones, Kang \&
Tregillis (1994) showed that among the knots interacting with the low
density ambient medium, those with a large density contrast parameter
$\chi$ ($\equiv\,\rho_{knot}/\rho_{amb}$) remain intact and
undecelerated longer than knots with smaller contrast parameters.
Anderson \etal~(1994) found that reverse-shocked knots follow roughly
a three-phase evolutionary process: (1) a bow shock phase, where the
knot drives a shock into the low density medium ahead of it and in the
process is itself compressed by a reverse shock, (2) an instability
phase where the reverse shock exits the back side of the knot and a
series of Rayleigh-Taylor and Kelvin-Helmholtz instabilities are
initiated, and (3) a dispersal phase where the knot is strongly
decelerated and the instabilities shred the structure into smaller
pieces.  Anderson \etal~(1994) noted that in a given knot the optical
emission from radiative shocks will cease once the shock emerges from
the rear of the knot.

The FMKs seen in [O~III] have not been significantly decelerated 
(see \S\ref{FMKs}), so we can safely conclude that these knots
have not yet reached the third (dispersal) phase of evolution.
On the other hand, Jones, Kang \& Rudnick (1994) predict that the
nonthermal radio emission from a knot will rise sharply during the
second (instability) phase, as shearing forces amplify magnetic fields
near the boundary of the knot.  Comparing the [O~III] image with the
20 cm continuum image of GW03, we find that some
FMK groups, such as the two northernmost groups seen in
Figure~\ref{g292_oiii+allaps}, are correlated with obvious patches of
enhanced radio emission (Figure~\ref{g292fmks_20cm+oiii}).  In contrast,
other FMKs, such as the southernmost group of elongated knots in
Figure~\ref{g292_oiii+allaps}, show little associated radio continuum
emission (Figure~\ref{g292fmks_20cm+oiii}).

The presence of both optical and radio emission from the FMKs suggests
that the knots are somewhere between the bow-shock and instability
phases of evolution. We note that the northernmost knots should be
further into the instability phase than the southern knots, because of
the greater relative prominence of radio emission from the northern
knots. Again, a higher density contrast for the southern knots would
be consistent with this scenario.  Since it is rare to detect X-ray,
optical, and radio emission from isolated fragments of ejecta in young
SNRs, further multiwavelength studies are clearly warranted.  However,
these are beyond the scope of the current article and are left to future
work.

{\it The Spur $-$} The most coherent ejecta structure seen in [O~III]
images of \g292\, shows little morphological correlation with features
detected in the \chan\, image.  As shown in the middle panel of
Figure \ref{g292opticalxray}, the top portion of the crescent-shaped
spur is seen at zero velocity, at the eastern end of the equatorial
bar structure.  The top of the spur does not match any bright
features in the \chan\, image.  The lower portion of the spur (the
redshifted structure seen out to +1700 \kms\, in Figure
\ref{g292redblue}) also shows little obvious X-ray emission.  This
suggests that virtually all of the main spur is composed of dense
O-rich material.  However, the streamers of material running southward
of the spur exhibit [O~III] clumps that lie at tips of X-ray knots.
This may indicate that the ejecta material in the streamers
is of more variable density, similar to what is seen in the FMKs.

\subsection{COMPARING THE OPTICAL AND X-RAY EMISSION FROM THE EQUATORIAL BAR }

The equatorial X-ray bar is perhaps the most prominent feature in the
broadband X-ray image of \g292. When this feature was first discovered
in the \eins\, image, it was
unclear whether the emission was from ejecta or circumstellar
material.  After finding that the kinematics of the [O~III] emission
in 1E 0102.2$-$7219 could be explained by expansion of a distorted
ring of high velocity ejecta, Tuohy \& Dopita (1983) suggested that
the equatorial belt seen in the \eins\, image of \g292\, could
be a similar structure seen nearly edge-on.  However, detailed
spectral analyses of the \chan\ X-ray data (Park \etal~2002, 2004;
Gonzalez \& Safi-Harb 2003) cast doubt on this interpretation when
these studies showed that the composition of the bar was generally
consistent with cosmic abundances.

\begin{figure}
\epsscale{1.}
\plotone{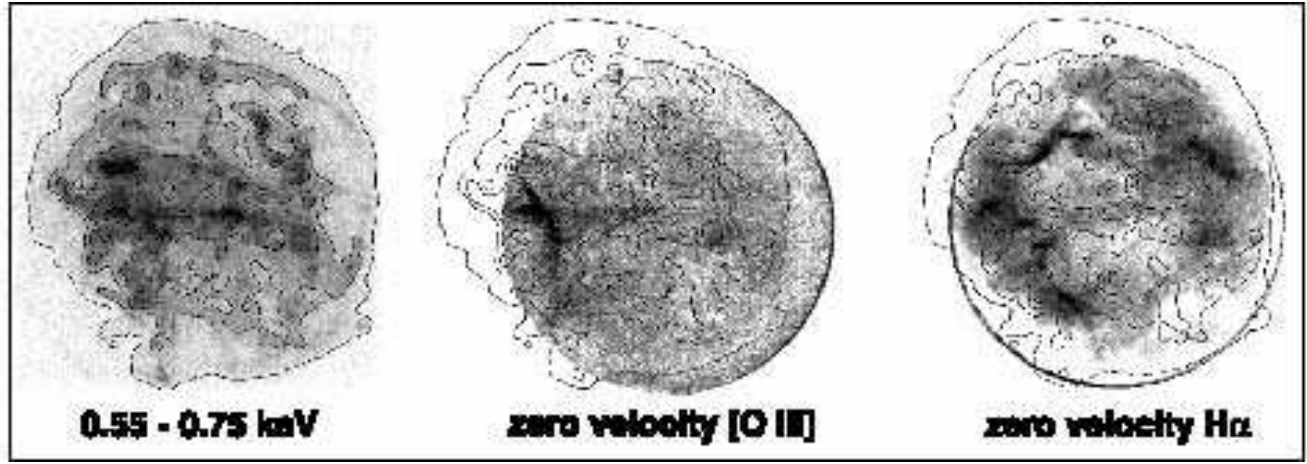}
\caption{The \chan\ X-ray image of \g292\ (shown far left) compared to
the [O~III] (middle panel) and H$\alpha$ (right panel) emission
detected by the RFP at zero heliocentric radial velocity.  Images are
shown to the same spatial scale.  The X-ray \chan\ image has been
filtered to show the oxygen line emission (0.55--0.75 keV) and has
been adaptively smoothed to bring out the fine structure. Contours of
the X-ray line emission are shown at logarithmic intervals of (0.05,
0.6, 1.07, 1.7, 2.6, 3.7 and 5.27)$\times$10$^{-4}$ cnts s$^{-1}$
arcsec$^{-2}$.  These contours are overlaid onto all three images. 
A full resolution
figure can be found at http://fuse.pha.jhu.edu/\~{}parviz/papers/g292 .
}
\label{g292opticalxray}
\end{figure}

\begin{figure}
\epsscale{0.8}
\plotone{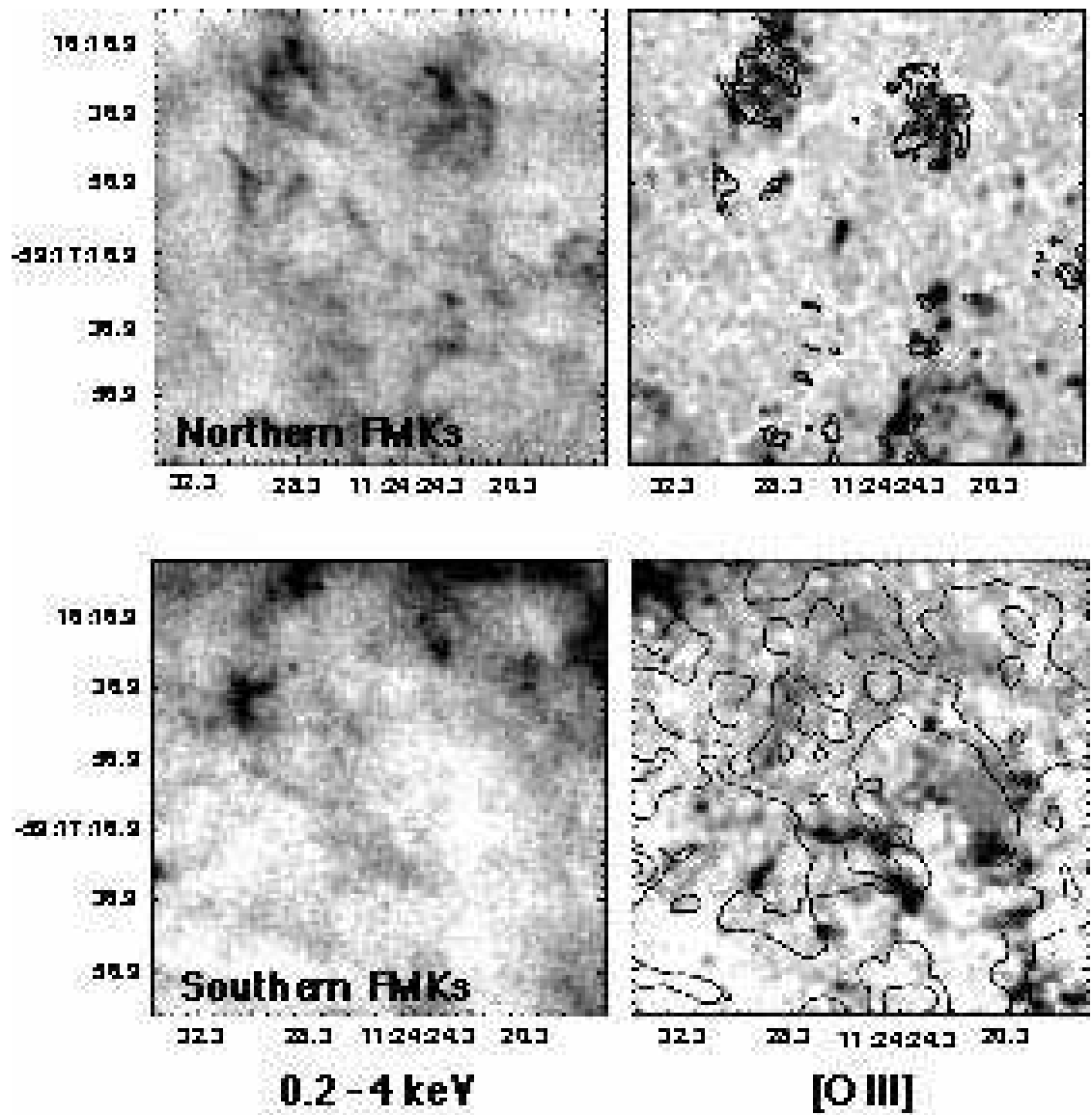}
\caption{
Closeup view of the northernmost and southernmost FMKs in
\g292, shown in the X-ray band (from \chan, left panels) and in the
narrowband [O III] image of WL05 (right panels).  The images are
displayed at the same spatial scale.  (The large circular gray patch
seen near the southern [O III] knots are the result of interpolation
over a bright star.) Contours of X-ray emission are overlaid onto the
optical image for comparison.  The contours of the northern FMKs are
spaced with a square root stretch at levels of (5.9, 9.5, 8.2 and
11.0)$\times$10$^{-4}$ cnts s$^{-1}$ arcsec$^{-2}$.  Contours of the
southern FMKs are also spaced with a square root stretch at levels of
3.0$\times$10$^{-4}$ and 4.3$\times$10$^{-4}$ cnts s$^{-1}$
arcsec$^{-2}$.   A full resolution
figure can be found at http://fuse.pha.jhu.edu/\~{}parviz/papers/g292 . }
\label{g292fmks_xray+oiii}
\end{figure}

\begin{figure}
\epsscale{0.7}
\plotone{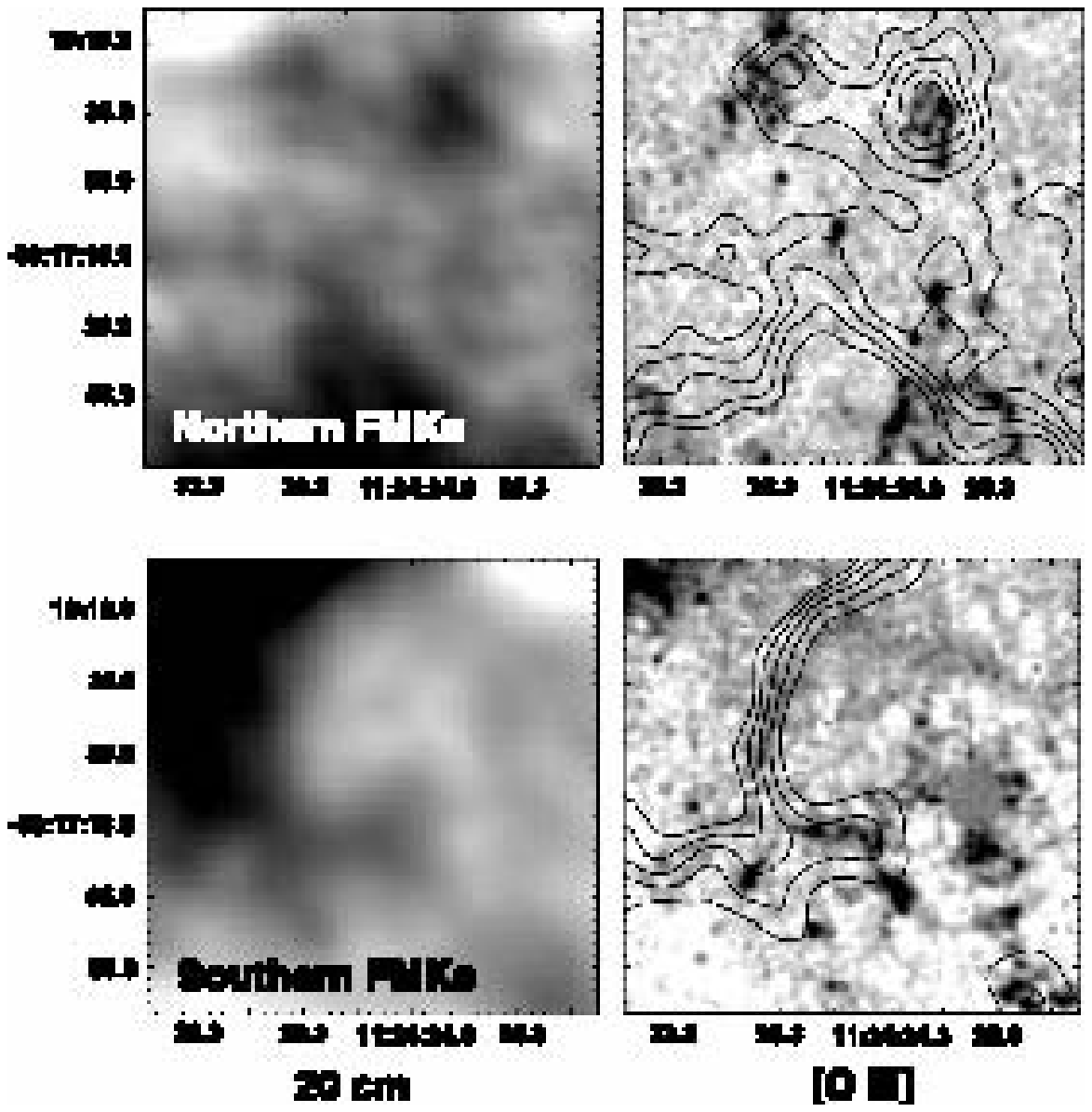}
\caption{
Closeup view of the northernmost and southernmost FMKs in
\g292, shown in the 20 cm radio continuum (GW03)
(left panels) and in the narrowband [O III] image of WL05 (right
panels).  The images are displayed at the same spatial scale and
correspond approximately to the same sky regions as
fig.~\ref{g292fmks_xray+oiii}.  Contours of the radio continuum are
overlaid onto the optical image for comparison.  The contours of the
northern FMKs are spaced with a logarithmic stretch at levels of 3.5,
3.9, 4.0, 4.6 and 5.0 mJy beam$^{-1}$.  Contours of the southern FMKs
are also spaced logarithmically at levels of 6.5, 7.4, 8.0, 9.0 and
10.0 mJy beam$^{-1}$.   A full resolution
figure can be found at http://fuse.pha.jhu.edu/\~{}parviz/papers/g292 .}
\label{g292fmks_20cm+oiii}
\end{figure}

A comparison of the optical and X-ray images in
Figure \ref{g292opticalxray}\, indicates that the zero-velocity features
B1 and B2 making up the [O~III] bar are likely the optical
counterparts of some parts of the X-ray structure seen by Park
\etal~(2002).  The gap in the bar is closely
matched in both the [O~III] and X-ray images.  In addition to the
belt, a circular patch of [O~III] emission 35\arcsec\, across can be
seen approximately 1\farcm3 southwest of B1 (feature P1
in Figure \ref{g292_oiii+allaps}).  In the X-ray image P1 lies at the
end of a ridge of material extending out from the center toward the
southwest, suggesting that they are part of the same structure.  Some
features such as B3 (Figure \ref{g292_markbelt}) are more difficult to
interpret since they appear to be circumstellar material (are seen
only at zero radial velocity), but exhibit no clearly associated
H$\alpha$ or X-ray emission.

Other than these cases, finding direct matches between H$\alpha$ and
X-ray features in \g292\ is less straightforward. The X-ray emission and zero
velocity H$\alpha$ emission in Figure \ref{g292opticalxray}\, show
little spatial correlation.  As mentioned earlier, there is no
X-ray emission from the H$\alpha$ filaments on the eastern side of the
remnant, indicating that they are not fast, non-radiative shocks in
neutral gas (i.e., Balmer-dominated shocks, Chevalier \& Kirshner
1978, Chevalier, Kirshner \& Raymond 1980).  

The most plausible explanation for the equatorial bar emission in
[O~III], H$\alpha$ and X-rays is that it is produced by partially
radiative shocks (i.e., shocks with incomplete cooling zones) in
moderate density ($n \sim 10$ cm$^{-3}$) circumstellar material.  The
incomplete cooling zones occur when the shock front has been moving 
through material for a time comparable to, but less than the cooling time
scale of the shocked gas.  Features B1, B2 and P1 in \g292\,
may be the sites of partially radiative shocks in cosmic abundance material.

Partially radiative shocks have been modeled for the case of the Cygnus Loop by Raymond
\etal~(1988), who showed that the relative flux of [O~III] to H$\beta$
(or H$\alpha$) is determined by the column density where the postshock
flow terminates.  Therefore, if the shock is viewed at a time before
the cooling zone has fully formed, it is possible for the output
[O~III] flux to significantly exceed that of both H$\beta$ and H$\alpha$.  A
truncated cooling zone in the equatorial bar would also explain its
overall optical faintness, since the emitting gas would not have
reached the full compression ($\sim\,$100) attainable in a fully
radiative shock.  The best way to confirm the existence of a partially
radiative shock would be to measure such ratios as [O~III]:H$\beta$, [O~II]/H$\beta$
and [O~I]:H$\alpha$ from a long-slit spectrum of the bar and to compare the
results with numerical shock model predictions.  This will be the
focus of a future study. 

\section{SUMMARY}

Clearly, \g292\, exhibits a very complex morphology. The optical study
of WL05 and our RFP kinematic investigation have only begun to reveal the rich
variations in elemental abundances, densities and kinematic properties
across this OSNR.  We believe that further study will ultimately prove
\g292\, to be every bit as interesting as Cas A.  From Rutgers
Fabry-Perot data we have extracted and fit [O~III] $\lambda$5007 line
profiles to 524 regions in \g292.  Our main results are as follows:

1.  The detection of numerous fast-moving knots of O-rich ejecta.
These FMKs are similar to the localized knots of ejecta seen in Cas A.
An interesting property of the FMKs is that very few are detected at
redshifted velocities; either the O-rich ejecta are preferentially
distributed on the near side of \g292\, or the reverse shock has not
yet encountered all the ejecta on the far side of the remnant.  The
position-velocity diagram of the FMKs exhibits the unmistakable
kinematic signature of an expanding shell expanding at a radial
velocity of 1700 \kms.  Assuming a distance of 6 kpc,
we derive an expansion age of 3000--3400
years for \g292.  Our fits indicate that the centroid of the FMK
velocity distribution is only slightly redshifted ($\sim$+100 \kms).
This systemic shift is substantially smaller than the shifts seen
in the other known OSNRs, and may reflect a difference in
explosion asymmetry between \g292\, and other OSNRs.

2. The bright spur of O-rich material which dominates the optical
appearance of \g292\, is kinematically resolved into a rippled shell
of predominantly redshifted ejecta.  However, interpreting the
position-velocity diagram of the spur is not straightforward.
Extracting spectra from a grid of regions covering the face of the
spur, we find that some points in the position-velocity diagram tend
to aggregate into groups of constant velocity.  These features are
seen at $-$300 \kms, 0 \kms\, and +950 \kms, and we interpret them as
ejecta clumps that have broken away from the main redshifted shell
of expanding material.  Morphologically, the wiggles seen
in the spur material at high redshifted velocities ($\geq$600 \kms)
are similar to the Rayleigh-Taylor unstable structures seen in Cas A.
As in Cas A, the wiggles in the spur of \g292\, point away from the
expansion center, indicating that the instabilities arise in
reverse-shocked ejecta.  This argues against the possibility that the
spur is ejecta swept from the inside out by the relativistic wind
nebula of the nearby pulsar PSR J1124$-$5916.

3. Optical detection of an equatorial bar matching portions of
the bright X-ray bar structure seen in prior \eins\, and
\chan\ images of \g292.  The bar is detected at the systemic radial
velocity of \g292 (0 \kms, or V$_{LSR}\,=\,$4.5 \kms\,) in both
[O~III] and H$\alpha$.  The optical emission is faint, suggesting that
the shocks driven into this material are just now becoming radiative.
There are several other structures near the bar, but not aligned with
it, which exhibit faint [O~III] and H$\alpha$ but are bright in
X-rays.  Our detection of the bar in both [O~III] and H$\alpha$
confirms the assertion by Park \etal~(2002, 2004) and Gonzalez \& Safi
Harb (2003) that this material is of cosmic composition.  The bar may
be portion of a ring of circumstellar material similar to what is seen in SN
1987A, but with the difference that it is nearly an order of magnitude
larger and of much lower density.  There are a number of FMKs along
the line of sight to the belt that are kinematically resolved in our
RFP scans but which give a stranded, chaotic appearance to this region
in the narrowband [O~III] images (WL05).

There appears to be an absence of fully radiative shocks in cosmic
composition material in \g292.  All observed shocks are either fully
non-radiative or partially radiative.  From our RFP spectra we are
unable to determine with certainty whether the belt is enhanced in
nitrogen as may be expected for a circumstellar wind.  Deep long-slit
spectroscopy and detailed modeling will be required to settle
this issue.

P.~G.\ would like to thank John Raymond for helpful discussions on the
shock physics presented in this paper, P.~F.\ Winkler for permission
to use the narrowband [O~III] image of \g292\, and Bryan Gaensler for
making the 20 cm radio image available.  The authors would also like
to thank the referee for very helpful suggestions on clarifying the
science of the paper and improving the presentation.  This work was
partially supported by Chandra Grants GO0-1035X and GO1-2052X, and NSF
grant AST 9619510.  J.~P.~H.\ and T.~B.~W.\ would like to thank the
staff of CTIO for their hospitality and support during the Fabry-Perot
observations.

\clearpage

\begin{references}

\reference{A94}
Anderson, M. C., Jones, T. W., Rudnick, L., Tregillis, I. L. \& Kang, H. 1994, \apj, 421, L31

\reference{BH78}
Balick, B. \& Heckman, T. 1978, \apj, 226, L7

\reference{bkw83}
Blair, W. P., Kirshner, R. P. \& Winkler, R. P. 1983, \apj\, 279, 708

\reference{B00}
Blair, W. P., \etal\, 2000, \apj, 537, 667

\reference{B83}
Braun, R., Goss, W. M., Danziger, I. J. \& Boksenberg, A. 1983, IAUS 101, 159

\reference{B86}
Braun, R., Goss, W. M., Caswell, J. L. \& Roger, R. S. 1986, \aap, 162, 259

\reference{C02}
Camilo, F., \etal\, 2002, \apj, 567, L71

\reference{C78}
Chevalier, R. A. \& Kirshner, R. P. 1978, \apj, 219, 931

\reference{C79}
--------- 1979, \apj, 233, 154

\reference{CKR80}
Chevalier, R. A., Kirshner, R. P. \& Raymond, J. C. 1980, \apj, 235, 186

\reference{C05}
Chevalier, R. A. 2005, \apj, 619, 839

\reference{C80}
Clark, D. H., Tuohy, I. R. \& Becker, R. H. 1980, \mnras, 193, 129

\reference{D81}
Dopita, M. A., Tuohy, I. R. \& Mathewson, D. S. 1981, \apj\, 248, L105

\reference{FBB87}
Fesen, R. A., Becker, R. H. \& Blair, W. P. 1987, \apj, 313, 378

\reference{F01}
Fesen, R. A. \etal, 2001, \aj, 122, 2644

\reference{G03}
Gaensler, B. M. \& Wallace, B. J. 2003 (GW03), \apj, 594, 326

\reference{G00}
Ghavamian, P., Raymond, J., Hartigan, P. \& Blair, W. P. 2000, \apj, 535, 266

\reference{GS03}
Gonzalez, M. \& Safi-Harb, S., 2003, \apj, 583, L94

\reference{G79}
Goss, W. M., Shaver, P. A., Zealey, W. J. Murdin, P. \& Clark, D. H. 1979,
\mnras, 188, 357

\reference{HRH87}
Hartigan, P. M., Raymond, J. C. \& Hartmann, L. 1987, \apj, 316, 323

\reference{HS94}
Hughes, J. P. \& Singh, K. P. 1994, \apj, 422, 126

\reference{H01}
Hughes, J. P., \etal\, 2001, \apj, 559, L153

\reference{H03}
Hughes, J. P., \etal\, 2003, \apj, 591, L139

\reference{JKT94}
Jones, T. W., Kang, H. \& Tregillis, I. L. 1994, \apj, 432, 194

\reference{K89}
Kirshner, R. P., Morse, J. A., Winkler, P. F. \& Blair, W. P. 1989, \apj, 342, 260

\reference{KMC94}
Klein, R. I., McKee, C. F. \& Colella, P. 1994, \apj, 420, 213

\reference{L78}
Lasker, B. M. 1978, \apj\, 223, 109

\reference{L80}
Lasker, B. M. 1980, \apj\, 237, 765

\reference{L95}
Lawrence, S. S. \etal, 1995, \aj, 109, 2635

\reference{L77}
Lockhart, I. A., Goss, W. M., Caswell, J. L. \& McAdam, W. B. 1977,
\mnras, 179, 147

\reference{M61}
Mills, B. Y., Slee, O. B. \& Hill, E. R. 1961, Au. J. Phys. 279, 708

\reference{M57}
Minkowski, R., 1957, in IAU Symp. 4, Radio Astronomy, ed. H .C. van de Hulst
(Cambridge: Cambridge Univ. Press), 107

\reference{M95}
Morse, J. A., Winkler, P. F. \& Kirshner, R. P. 1995, \aj, 109, 2104

\reference{M96}
Morse, J. A., \etal\, 1996, \aj, 11, 509

\reference{M79}
Murdin, P. \& Clark, D. H. 1979, \mnras\ 189, 501

\reference{P02}
Park, S., \etal\, 2002, \apj, 564, L39

\reference{P04}
Park, S., \etal\, 2004, \apj\, 602, L33

\reference{R88}
Raymond, J. C., Hester, J. J., Cox, D., Blair, W. P., Fesen, R. A. \& Gull,
T. R. 1988, \apj, 324, 869

\reference{R95}
Reed, J. E., Hester, J. J., Fabian, A. C. \& Winkler, P. F. 1995, \apj, 440, 707

\reference{R85}
Reynolds, S. P. 1985, \apj\, 291, 152

\reference{S04}
Serafimovich, N. I., Shibanov, Yu. A., Lundqvist, P. \& Sollerman, J. 2004, \aap, 425, 1041

\reference{SD95}
Sutherland, R. S. \& Dopita, M. A. 1995, \apj, 439, 365

\reference{T82}
Tuohy, I. R., Clark, D. H. \& Burton, W. M. 1982, \apj, 260, 65

\reference{T83}
Tuohy, I. R. \& Dopita, M. A. 1983, \apj\, 268, L11

\reference{v79}
van den Bergh, S. 1979, \apj, 234, 493

\reference{W85}
Winkler, P. F. \& Kirshner, R. P. 1985, \apj\, 299, 981

\reference{WL05}
Winkler, P. F. \& Long, K. S. 2005, \apj, in preparation (WL05)


\end{references}
\end{document}